\documentclass[aps,pra,twocolumn,nofootinbib,superscriptaddress]{revtex4-1}

\usepackage{amsmath, amssymb, graphicx}
\usepackage[american]{babel}

\usepackage[section]{placeins}
\usepackage{graphicx}
\usepackage{dcolumn}
\usepackage{bm}

\usepackage{mathrsfs}

\usepackage{verbatim}

\usepackage{comment}

\usepackage[usenames]{color}

\newcommand{\be}{\begin{equation}}
\newcommand{\ee}{\end{equation}}
\def\beq#1\eeq{\begin{align}#1\end{align}}
\newcommand{\beeq}{\begin{eqnarray}}
\newcommand{\eeeq}{\end{eqnarray}}

\def\H1{\widehat{H}_1}

\newcommand{\pd}{\partial}

\newcommand{\lb}{\left[}
\newcommand{\rb}{\right]}
\newcommand{\lp}{\left(}
\newcommand{\rp}{\right)}

\def\op#1{{\Hat{\mathrm{#1}}}}

\def\bra#1{\ensuremath{\langle{#1}\vert}}
\def\ket#1{\ensuremath{\vert{#1}\rangle}}
\def\bracket#1#2{\ensuremath{%
    \langle{#1}\mkern1.2mu\vert\mkern1.2mu{#2}\rangle}}

\def\bbra#1{\ensuremath{\langle\kern-0.2em\langle{#1}\vert}}
\def\kket#1{\ensuremath{\vert{#1}\rangle\kern-0.2em\rangle}}
\def\bbrackket#1#2{\ensuremath{%
    \langle\kern-0.2em\langle{#1}\mkern1.2mu\vert\mkern1.2mu{#2}\rangle\kern-0.2em\rangle}}

\def\commutator#1#2{\mathinner{%
    \mathopen[#1,#2\mathclose]}}

\def\anticommutator#1#2{\mathinner{%
    \mathopen\{#1,#2\mathclose\}}}

\def\abs#1{\mathinner{\lvert#1\rvert}}
\def\norm#1{\mathinner{\lVert#1\rVert}}

\newcommand{\1}{\leavevmode{\rm 1\ifmmode\mkern  -4.8mu\else\kern -.3em\fi I}}

\begin{document}



\title{Accuracy of the adiabatic-impulse approximation for closed and open quantum systems}




\author{Michael Tomka}
\email[]{mtomka@usc.edu}
\affiliation{Department of Physics \& Astronomy, University of Southern California, Los Angeles, California 90089, USA}
\affiliation{Center for Quantum Information Science \& Technology, University of Southern California, Los Angeles, California 90089, USA}
\author{Lorenzo Campos Venuti}
\affiliation{Department of Physics \& Astronomy, University of Southern California, Los Angeles, California 90089, USA}
\affiliation{Center for Quantum Information Science \& Technology, University of Southern California, Los Angeles, California 90089, USA}
\author{Paolo Zanardi}
\affiliation{Department of Physics \& Astronomy, University of Southern California, Los Angeles, California 90089, USA}
\affiliation{Center for Quantum Information Science \& Technology, University of Southern California, Los Angeles, California 90089, USA}

\date{December 12, 2017}


\begin{abstract}
We study the adiabatic-impulse approximation (AIA) as a tool to approximate
the time evolution of quantum states, when driven through a region of
small gap.
The AIA originates from the Kibble-Zurek
theory applied to continuous quantum phase transitions.
The Kibble-Zurek mechanism was developed to predict the power-law scaling of the
defect density across a continuous quantum phase transition.
Instead here, we quantify the accuracy of the AIA via the trace norm
distance with respect to the exact evolved state.
As expected, we find that for short times/fast protocols, the AIA
outperforms the simple adiabatic approximation.
However, for large times/slow protocols, the situation is actually
reversed and the AIA provides a worse approximation.
Nevertheless, we found a variation of the AIA that can perform better
than the adiabatic one. This counter-intuitive modification consists
in crossing {\it twice} the region of small gap.
Our findings are illustrated by several examples of driven closed and
open quantum systems.
\end{abstract}


\maketitle



\section{Introduction}
\label{sec:introduction}

Progress made during the last thirty years in the field of atomic and
molecular optics, in experiments with trapped ions, and in cavity and
circuit quantum electrodynamics, has drastically improved the
experimental control over the dynamics of quantum many-body systems.
These experimental implementations of controllable quantum
systems~\cite{Bloch2005, Bloch2008, Bloch2012},  
opened the possibility to use quantum physics towards the realization
of quantum technologies like quantum computers~\cite{Lloyd2000, Farhi2001, NielsenChuang2000} 
and quantum simulators~\cite{Feynman1982}.
Among the different approaches to quantum computing, the adiabatic one
is recently attracting a lot of attention~\cite{Denchev2016, Boixo2016, Lanting2014, Dickson2013, Johnson2011, Albash2017}.
The basic idea behind adiabatic quantum computation is that
the ground-state of certain quantum systems can encode the solution to
a mathematical problem, e.g., the solution of a minimization problem.
The algorithm is to start with a simple Hamiltonian whose ground-state
can easily be prepared. 
In order to get from this easy available ground-state, to the target
ground-state, encoding the solution of the minimization problem, one
adiabatically evolves the simple Hamiltonian to the desired
complicated Hamiltonian.
According to the adiabatic theorem, the system remains in the
same level, if the total evolution time is large enough, such that
the system ends up being in the state describing the solution of the
minimization problem.

It is clearly very important to understand the precise mode of
operation of such an adiabatic quantum algorithm, in order to obtain
faithful results and to understand the limit of its performance.
Key problems are controlling the precision of the initial ground-state
preparation, having full control over the system's parameters and
understanding the main features that control the adiabatic evolution
of the quantum many-body system, i.e., being aware of when the energy
gap, the energy difference between the ground-state and the first
excited-state, becomes small, as well as the effects of dissipation
and decoherence.
Examples where the adiabatic dynamics can be analyzed in full details
are rare and only possible for very small quantum systems, therefore
to understand and fully quantify the performance of adiabatic quantum
computers one needs to relay on approximation methods.

Consequently, in the present work we study the adiabatic-impulse
approximation (AIA) to estimate the time evolution of quantum states.
The idea behind the AIA is that the time evolution can approximately
be divided in an adiabatic and an impulse stage (the impulse
stage is sometimes also called sudden or diabatic stage).
During the adiabatic stage the external changes are slow
compared to the {\it internal time scale} of the system, such that the
adiabatic approximation becomes appropriate.
Conversely, in the impulse region the external changes happen so fast that
the state has no time to adjust itself, and the impulse approximation
is a good one.
The difficulty of the AIA lies in the determination of the precise
internal time scale of the problem, and/or in the identification
of the switching instants: adiabatic to impulse and vice versa.
Hence, the paradigmatic situation where the AIA can be applied,
appears when the system is driven across a quantum critical point.

Damski~\cite{Damski2005} applied the AIA to study the quantum dynamics
of the excitations in the Landau-Zener model.
It was pointed out, that the AIA is based on the Kibble-Zurek (KZ)
theory of non-equilibrium classical phase
transitions~\cite{Damski2005, Damski2006, Dziarmaga2010}.
The KZ theory provides one way to determine this internal time scale,
namely, assuming that it is given by the inverse gap.
This recipe fixes the time scale apart from a dimensionless constant,
that traditionally is fixed by comparing the approximation of the
density of excitation to the analytical expression~\cite{Damski2005, Damski2006, Dziarmaga2010}.
In this paper we will carefully examine different strategies to fix
this internal time scale, which allows us to estimate
the adiabatic-impulse switching times.
The accuracy of the resulting AIA is evaluated by considering the
trace norm distance between the obtained approximation and the
numerically performed exact evolution. 

The scaling prediction of the KZ mechanism have been confirmed in a
series of works~\cite{KZE}.
However we note, that the same scaling predictions can be obtained
without resorting to the AIA~\cite{Polkovnikov2005, Polkovnikov2010}.

The paper is organized as follows.
Section~\ref{sec:aiaclosed} gives a short review of the AIA in closed systems.
Then it is applied to approximate the time evolution of two
paradigmatic examples, namely the Landau-Zener (LZ) model and the
transverse field Ising (TFI) model.
The AIA method is evaluated by studying the distance between the
exact evolved state, which is computed numerically, and the one
obtained by the AIA.
In Sec.~\ref{sec:aiaopen} we will extend the AIA to
approximate the time evolution of open quantum systems.
More specifically, we consider a dissipative quantum system, where the
dynamics are described by a time-dependent Lindblad master-equation in
the Davies form.
As an example, we consider a single qubit coupled to a thermal bath
and study the AIA as in the closed case.
A brief summary is presented in the concluding Sec.~\ref{sec:conclusions}.
Appendixes~\ref{appendix:A} and~\ref{appendix:B}
give some details on the adiabatic intertwiner that evolves the states
corresponding to the eigenvector of the Liouvillian with zero
eigenvalue and the full adiabatic intertwiner that evolves all the
eigenvectors together, respectively.
In Appendix~\ref{appendix:C} we derive the eigenvalues and
eigenvectors of the Liouvillian describing the single qubit coupled to
a thermal bath and Appendix~\ref{appendix:D} shows the corresponding
evolution equations.

\section{Adiabatic-impulse approximation in closed systems}
\label{sec:aiaclosed}

In this section, we examine the accuracy of the adiabatic-impulse
approximation (AIA) method for the time evolution of isolated quantum
systems, that are driven through a region of minimal gap.
First, we will review the basic ideas of the AIA for closed systems.
We evaluate the AIA by computing the distance between the fully
evolved state, obtained by numerically propagating the time-dependent
Schr\"odinger equation, and the state obtained by the AIA. 
As a comparison we use the simple adiabatic approximation, and consider the
distance between the fully evolved state and the adiabatic approximation. 
This will be illustrated by the examples of the Landau-Zener model
(avoided level crossing) and the transverse field Ising model (quantum
phase transition).

\subsection{General Setting}
\label{subsec:generalseting}

Let us consider a closed quantum system described by a time-dependent
Hamiltonian $\op{H}(t)$, whose instantaneous
eigenstates and eigenenergies are and given by
\be
\op{H}(t)\ket{\psi_{n}(t)} = E_{n}(t) \ket{\psi_{n}(t)},
\ee
with $n=1,2,\ldots, \dim{\mathcal{H}}$, where $\dim{\mathcal{H}}$ is
the dimension of the Hilbert space $\mathcal{H}$.
Just for simplicity we consider the Hamiltonian to be non-degenerate.
We label the ground-state by $n=1$, the first-excited state by $n=2$,
and so on.
Further, we assume that the time-dependence enters through a
single parameter denoted by $\lambda(t)$.

We focus on dynamics that include both an adiabatic and an impulse
regime, e.g., the crossing of a quantum critical point.
The unitary time evolution of a closed quantum system is adiabatic,
when the system initialized in an eigenstate
$\ket{\psi_{m}(t_{i})}$ will remain in it $\ket{\psi_{m}(t)}$ for all
$t \in [t_{i},t_{f}]$, where $t_{i}$ and $t_{f}$ denote the initial
and the final time, respectively.
A ``folklore'' condition that the evolution is adiabatic can be given by
\be
\max_{t\in[t_{i},t_{f}]} 
\frac{\abs{\bra{\psi_{n}} \pd_{t} \op{H} \ket{\psi_{m}}}}
     {\abs{E_{n}-E_{m}}} \ll \min_{t\in[t_{i},t_{f}]} \abs{E_{n}-E_{m}}, 
\quad \forall \, n \neq m,
\ee
see~\cite{Messiah1962}.
In the region where the gap becomes minimal,
the time evolution becomes diabatic (impulse regime).
During the impulse regime the system can no longer adjusts to the
external changes in the Hamiltonian and therefore its state is
effectively frozen.
The time evolution of the wave-function is thus approximated by a
sudden jump through this regime, in other words, no changes in the
wave-function occur.

Our protocol will be the following, we initialize the system at $t_{i}=0$
in the ground-state $\ket{\psi_{0}(0)}$ and then tune the parameter
$\lambda(t)$ from its initial value $\lambda_{i}=\lambda(0)$ 
to its final value $\lambda_{f}=\lambda(t_{f})$.
We assume that the gap, $\Delta \equiv E_{1}-E_{0}$, will be minimal at a
single instant in time.
Within the AIA the evolution is assumed to
be adiabatic until the instant $\tau_{-}$ and again adiabatic
after $\tau_{+}$ and the minimum of the gap occurs within the interval
$[\tau_{-},\tau_{+}]$.
During the interval $[\tau_{-},\tau_{+}]$ the state of the system is
assumed not change, it suddenly jumps from $\tau_{-}$ to $\tau_{+}$.


%
The Kibble-Zurek argument used
in~\cite{Damski2005, Damski2006, Dziarmaga2010} presumes
the impulse instants $\tau_{\pm}$ to be determined by the time, when
the transition time, $\abs{\frac{\lambda}{\pd_{t}\lambda}}$, is equal
to the inverse gap, $1/\Delta$, 
\be
\left|\frac{\lambda(t)}{\pd_{t}\lambda(t)}\right|_{t=\tau}
=
\frac{1}{\Delta(\lambda(\tau))}.
\label{eq:eqtcone}
\ee
This equation is the adaptation from the so-called Kibble-Zurek theory
of topological defect production during classical phase
transition~\cite{Kibble1976, Kibble1980, Zurek1985, Zurek1993, Zurek1996},
where the corresponding crossover time is determined by the condition 
$t_{\mathrm{rel}}(\tau)=\tau$, $t_{\mathrm{rel}}$ being the relaxation
time scale of the system.
In order to adapt the KZ theory to quantum systems, the identification
$t_{\mathrm{rel}}=1/\Delta$ was
made in~\cite{Damski2005}, to obtain Eq.~(\ref{eq:eqtcone}).
Within the following examples, the Landau-Zener model and the transverse
field Ising model, we will examine, if the
condition~(\ref{eq:eqtcone}) faithfully estimates the impulse instants
or if one needs to find a more refined estimate to improve the
AIA.

The fully time evolved state, $\ket{\psi(t)}$, is given by the
solution of the Schr\"odinger equation,
$\pd_{t}\ket{\psi(t)} = -i \, \op{H}(t) \ket{\psi(t)}$, 
which can formally be written as
\be
\ket{\psi(t_{f})}
=
\overleftarrow{\mathrm{T}}\!\!\exp
 \lb
    \int_{0}^{t_{f}}dt (-i) \op{H}(t)
 \rb
\ket{\psi(0)},
\ee
where $\overleftarrow{\mathrm{T}}$ is the time-ordering operator,
which arranges operators in a chronological order with time increasing
from right to left.
We note, that for all the examples considered here we computed the
time evolution numerically.

The adiabatic approximation of the state $\ket{\psi(t_{f})}$ is given by
\begin{align}
\ket{\psi_{\mathrm{adi}}(t_{f})}
&=
\op{U}(t_{f},0)\ket{\psi(0)},
\end{align}
where 
$\op{U}(t_{f},0)
=
\sum_{n}
e^{i \phi_{n}(t_{f},0)}
\ket{\psi_{n}(t_{f})}\bra{\psi_{n}(0)}$,
is the full adiabatic intertwiner~\cite{Kato1950}, and 
$\phi_{n}(t_{f},0) = - \delta_{n}(t_{f},0) + \gamma_{n}(t_{f},0)$
is the sum of the dynamic phase of the $n$-th eigenstate
$\delta_{n}(t_{f},0) = \int_{0}^{t_{f}} E_{n}(t) \, dt$
and the corresponding geometric phase
$\gamma_{n}(t_{f},0) = \int_{0}^{t_{f}} i \, \bra{\psi_{n}}\pd_{t}\ket{\psi_{n}} \, dt$.
If the initial state is the ground-state
$\ket{\psi(0)}=\ket{\psi_{1}(0)}$, the last equation reduces to
\be
\ket{\psi_{\mathrm{adi}}(t_{f})}
=
e^{i\phi_{1}(t_{f},0)}
\ket{\psi_{1}(t_{f})}.
\ee

A more refined approximation might be given by the AIA, where the
evolution is adiabatic before $\tau_{-}$, $(0<t<\tau_{-})$, and again
after $\tau_{+}$, $(\tau_{+}<t<t_{f})$, 
but it suddenly jumps from $\tau_{-}$ to $\tau_{+}$. 
Consequently, within the AIA scheme the
time evolved state is approximated by
\be
\ket{\psi_{\mathrm{aia}}(t_{f})}
=
\op{U}(t_{f},\tau_{+})
\,
\op{1}
\,
\op{U}(\tau_{-},0)
\ket{\psi(0)},
\ee
and which for $\ket{\psi(0)}=\ket{\psi_{1}(0)}$ reduces to
\begin{align}
&\ket{\psi_{\mathrm{aia}}(t_{f})}=
\nonumber \\
&
\sum_{n}
e^{i \phi_{n}(t_{f},\tau_{+})}
e^{i \phi_{0}(\tau_{-},0)}
\bracket{\psi_{n}(\tau_{+})}{\psi_{1}(\tau_{-})}
\ket{\psi_{n}(t_{f})}.
\end{align}

As a measure to quantify the adiabatic approximation and the AIA, we
use the distance between two given wave-functions,
$\ket{\psi}\in\mathcal{H}$ and $\ket{\phi}\in\mathcal{H}$, 
\be
\label{eq:distclosed}
\mathrm{d}[\ket{\psi},\ket{\phi}]
=
\sqrt{1-\abs{\bracket{\psi}{\phi}}^2},
\ee
which we note is defined in terms of the fidelity
$\mathcal{F}(\ket{\psi},\ket{\phi})=\abs{\bracket{\psi}{\phi}}^{2}$.
The distance between the fully evolved state $\ket{\psi(t_{f})}$ and
the adiabatic approximation is denoted by 
$\mathrm{d}_{\mathrm{adi}}(t_{f})
=
\mathrm{d}[\ket{\psi(t_{f})},\ket{\psi_{\mathrm{adi}}(t_{f})}]$,
while the distance between the fully evolved state and the one
obtained by the AIA is labeled
$\mathrm{d}_{\mathrm{aia}}(t_{f})
=
\mathrm{d}[\ket{\psi(t_{f})}, \ket{\psi_{\mathrm{aia}}(t_{f})}]$.

\subsection{Landau-Zener model}
\label{subsec:lzm}

As a first example, we consider the Landau-Zener model, described by
the Hamiltonian
\be
\op{H}_{\mathrm{LZ}}(t)
=
x(t)\op{\sigma}^{x}
+
z(t)\op{\sigma}^{z}
=
\begin{pmatrix}
z(t) & x(t) \\
x(t) & -z(t)
\end{pmatrix},
\ee
where $\op{\sigma}^{x}$ and $\op{\sigma}^{z}$ are the usual Pauli
matrices, and $\ket{\varphi_{1}}=(1,0)^{T}$, $\ket{\varphi_{2}}=(0,1)^{T}$,
denote the eigenstates of $\op{\sigma}^{z}$.
The parameter $x$ characterizes the coupling between
the two levels and $z$ the detuning.
The eigenenergies of this system are $E_{1,2} = \mp b$, where we
defined $b\equiv\sqrt{x^{2}+z^{2}}$, and the corresponding eigenstates
read
\be
\label{eq:eigvecslz}
\ket{\psi_{1,2}(t)}
=
\mp \sqrt{\frac{b \mp z}{2\,b}} \ket{\varphi_{1}}
+   \sqrt{\frac{b \pm z}{2\,b}} \ket{\varphi_{2}}.
\ee
We assume the protocol, where $x$ is constant in time, $z(t)$ changes
linear, $z(t)=z_{i}+(z_{f}-z_{i})\,t/t_{f}$, with
$t \in \lb 0, t_{f} \rb$, and the system is initially prepared in
the ground-state, $\ket{\psi(0)}=\ket{\psi_{1}(0)}$. 
For the initial point, $z_{i}$, we choose a negative value, and for the
final point, $z_{f}$, a positive value, such that the protocol
passes the avoided level crossing at $z=0$.
Let us note, that our protocol is similar to the paradigmatic Landau-Zener
problem~\cite{Landau1932, Zener1932, Stueckelberg1932, Majorana1932},
however, in the Landau-Zener problem one has, $z(t)=t/t_{f}$, with
$t\in\lb-\infty,\infty\rb$.
The Schr\"odinger equation,
$\pd_{t}\ket{\psi}=-i\,\op{H}_{\mathrm{LZ}}\ket{\psi}$, 
written in the fixed basis,
$\ket{\psi(t)}=\sum_{i=1}^{2}c_{i}(t)\ket{\varphi_{i}}$,
becomes
\be
\label{eq:schroedeqlz}
i \, \pd_{t} c_{1} = z(t) \, c_{1} + x \, c_{2},
\qquad
i \, \pd_{t} c_{2} = x \, c_{1} - z(t) \, c_{2},
\ee
with the initial conditions given by the ground-state
\be
c_{1}(0)
=
-
\sqrt{\frac{b_{i}-z_{i}}{2 \, b_{i}}},
\qquad
c_{2}(0)
=
\sqrt{\frac{b_{i}+z_{i}}{2 \, b_{i}}},
\ee
where $b_{i}\equiv\sqrt{x^{2}+z_{i}^{2}}$.
This system can be solved in terms of parabolic cylinder
functions~\cite{Vitanov1996},
and therefore provides a convenient benchmark to study the accuracy of the
adiabatic-impulse approximation.

\begin{figure}
\includegraphics[scale=0.9,trim=0.05mm 0.05mm 0.05mm 0.05mm, clip]{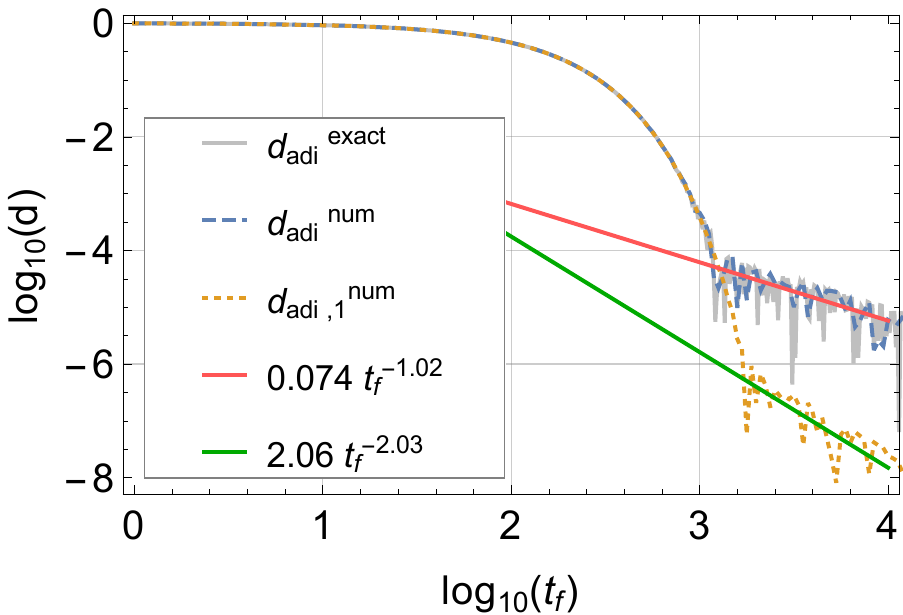}
\caption{
  (Color online)
  We plot $\mathrm{d}_{\mathrm{adi}}$ and $\mathrm{d}_{\mathrm{adi},1}$
  as a function of the total evolution time $t_{f}$ for $z_{i}=-1$,
  $z_{f}=1$ and $x=0.1$.
  The gray solid line corresponds to the distance between the
  adiabatic approximation $\ket{\psi_{\mathrm{adi}}(t_{f})}$ and the
  time evolved state $\ket{\psi(t_{f})}$ obtained by in terms of the
  parabolic cylinder functions, denoted by 
  $\mathrm{d}_{\mathrm{adi}}^{~~~\mathrm{exact}}$.
  To check our numerical procedure for solving the Schr\"odinger
  equation we plotted also
  $\mathrm{d}_{\mathrm{adi}}^{~~~\mathrm{num}}$ (blue dashed line),
  which corresponds to the distance between the adiabatic
  approximation $\ket{\psi_{\mathrm{adi}}(t_{f})}$ and the time
  evolved state $\ket{\psi(t_{f})}$ found by numerically solving 
  Eq.~(\ref{eq:schroedeqlz}).
  We see a perfect agreement between the exact and the numerically
  obtained distance and therefore use the same numerical procedure for
  the other examples in the text.
  The orange dotted line shows the distance between the first order
  correction to the adiabatic approximation and the fully time evolved
  state, $\mathrm{d}_{\mathrm{adi},1}$.
  }
\label{fig:lzdadi}
\end{figure}

We begin by studying the simple adiabatic approximation of the time
evolved state, $\ket{\psi(t_{f})}$, which is given by
\be
\label{eq:adiapproxslz}
\ket{\psi_{\mathrm{adi}}(t_{f})}
=
e^{-i \delta_{1}(0,t_{f})}
\ket{\psi_{1}(t_{f})},
\ee
and where the dynamical phase of the ground-state reads
\be
\delta_{1}(0,t_{f})
\equiv
\int_{0}^{t_{f}} E_{1}(t) \, dt
=
\frac{t_{f}}{2 \, \delta\!z}
\left.
\lb
b \, z + x^{2}\log(z+b)
\rb
\right|_{z_{i}}^{z_{f}},
\ee
with $\delta\!z \equiv z_{f}-z_{i}$.
Note that there is no Berry phase,
$\gamma_{1}(0,t_{f})
=
\int_{0}^{t_{f}} dt\, i \bra{\psi_{1}}\pd_{t}\ket{\psi_{1}}=0$,
since the Landau-Zener Hamiltonian is real.
In Fig.~\ref{fig:lzdadi} we plot $\mathrm{d}_{\mathrm{adi}}(t_{f})$ on
a logarithmic scale for $z_{i}=-1, z_{f}=1$ and $x=0.1$.
It can be seen that for large $t_{f}$ the distance decreases with
$t_{f}^{-1}$, as expected by the adiabatic theorem~\cite{Jansen2007}.

Next, we analyze the first-order correction of the adiabatic
approximation, which can be expressed by
\be
\label{eq:adifoclz}
\ket{\psi_{\mathrm{adi},1}(t_{f})}
=
N^{-1}
\lb
\ket{\psi_{\mathrm{adi}}^{\phantom{(0)}}(t_{f})}
+
\frac{1}{t_{f}}
\ket{\psi_{\mathrm{adi}}^{(1)}(t_{f})}
\rb,
\ee
where
\begin{align}
\ket{\psi_{\mathrm{adi}}^{(1)}(t_{f})}
&=
i \, e^{-i \delta_{1}(0,t_{f})} J_{21}(t_{f}) \ket{\psi_{1}(t_{f})}
\nonumber \\
&-
i \, e^{-i \delta_{1}(0,t_{f})} M_{21}(t_{f}) \ket{\psi_{2}(t_{f})}
\nonumber \\
&+
i \, e^{-i \delta_{2}(0,t_{f})} M_{21}(0) \ket{\psi_{2}(t_{f})},
\end{align}
and where we used the notations
\begin{align}
J_{21}(t)
&=
t_{f} \int_{0}^{t}
\frac{\abs{\bra{\psi_{2}}\pd_{t'}\op{H}_{\mathrm{LZ}}\ket{\psi_{1}}}^{2}}
     {(E_{2}-E_{1})^{3}} dt',
\\
M_{21}(t)
&=
t_{f}
\frac{\bra{\psi_{2}}\pd_{t}\op{H}_{\mathrm{LZ}}\ket{\psi_{1}}}
     {(E_{2}-E_{1})^{2}}.
\end{align}
The dynamical phase of the excited-state is given by
$\delta_{2}(0,t_{f}) = -\delta_{1}(0,t_{f})$, and the normalization
reads
$
N^{2}
=
1+\bracket{\psi_{\mathrm{adi}}^{(1)}(t_{f})}{\psi_{\mathrm{adi}}^{(1)}(t_{f})}$.
For a derivation of Eq.~(\ref{eq:adifoclz}) see
Ref.~\cite{Rigolin2008}.
The distance between the exactly evolved state and the first order
correction,
$
\mathrm{d}_{\mathrm{adi},1}(t_{f})
=
\mathrm{d}[\ket{\psi(t_{f})}, \ket{\psi_{\mathrm{adi},1}(t_{f})}]
$,
is also plotted in Fig.~\ref{fig:lzdadi}. 
Correctly, the distance $\mathrm{d}_{\mathrm{adi},1}(t_{f})$
decreases with, $2.06~t_{f}^{-2.03}$, for large $t_{f}$, and hence gets
much smaller than $\mathrm{d}_{\mathrm{adi}}(t_{f})$.

Let us now turn to the AIA
and study how its accuracy compares to the adiabatic expansion.
The time evolved state within the AIA for our Landau-Zener model
reduces to
\begin{align}
&\ket{\psi_{\mathrm{aia}}(t_{f})}
=
\nonumber \\
&\sum_{j=1}^{2}
e^{-i \delta_{j}(\tau_{+},t_{f})} e^{i \delta_{1}(0,\tau_{-})}
\bracket{\psi_{j}(\tau_{+})}{\psi_{1}(\tau_{-})}
\ket{\psi_{j}(t_{f})},
\end{align}
where the Berry phase of the excited-state is also zero, due to the
fact that the Hamiltonian is real.
As mentioned above, the difficulty of the AIA is the determination of
the time instants $\tau_{-}$ and $\tau_{+}$, where the evolution
switches from adiabatic to impulse and back from the impulse regime to
adiabatic, respectively.
In the following we will discuss different scenarios providing
the instants $\tau_{\pm}$.

\subsubsection{Switching instants $\tau_{1,\pm}$: standard Kibble-Zurek argument}

First, we consider the Kibble-Zurek argument, as proposed
in~\cite{Damski2005, Damski2006, Dziarmaga2010}.
The argument is based on the heuristic, that sufficiently
close to the critical point, here the avoided level crossing, the
dynamics appears to be ``frozen''.
The system's dynamics has not enough time to adjust to the changes of
the external parameters, due to the smallness of the gap 
$\Delta \equiv E_{2}-E_{1} = 2 b$.
According to the Kibble-Zurek argument this critical slowing down
occurs, when the inverse of the gap is on the order of the inverse rate of
change of the external parameter, therefore $\tau_{\pm}$ are
determined by
\be
\label{eq:kzatau}
\frac{1}{\Delta(\tau)} = \left| \frac{z(t)}{\pd_{t}z(t)}\right|_{t=\tau}.
\ee
This equation has two real solutions
\be
\tilde{\tau}_{1,\pm}
=
-
\frac{z_{i}}{\delta\!z} t_{f}
\pm
\frac{x}{\sqrt{2} \, \delta\!z} t_{f}
\sqrt{-1+\sqrt{1+\lp\frac{\delta\!z}{x^{2}} \frac{1}{t_{f}}\rp^{2}}}.
\ee
We note that the instants $\tau_{\pm}$ have to be positive and smaller
than or equal to the total evolution time $t_{f}$, since for our
protocol $t \in \lb 0, t_{f} \rb$, and therefore we find
\be
\tau_{1,\pm}
=
\begin{cases}
  \begin{cases}
    t_{f}
    \\
    0
  \end{cases},
  &
  \qquad
  0<t_{f}<\frac{1}{2}\frac{\delta\!z}{z_{f} \sqrt{x^{2}+z_{f}^{2}}}
  \\
  \tilde{\tau}_{1,\pm},
  &
  \qquad
  \frac{1}{2}\frac{\delta\!z}{z_{f} \sqrt{x^{2}+z_{f}^{2}}}<t_{f}<\infty
\end{cases}.
\ee
In Fig.~\ref{fig:lzdeltatau} we plot the resulting impulse interval,
$\Delta\tau_{1}\equiv\tau_{1,+}-\tau_{1,-}$, as
a function of $t_{f}$ for $z_{i}=-1$, $z_{f}=1$ and $x=0.1$
(red dotted line).
We note, the interval $\Delta\tau_{1}$ reaches the constant
value $1/x$ in the limit of $t_{f}$ approaching infinity.
The distance between the AIA
$\ket{\psi_{\mathrm{aia}-1}(t_{f})}$ and the fully
evolved state (numerical solution of Eq.~(\ref{eq:schroedeqlz})),
$
\mathrm{d}_{\mathrm{aia}-1}(t_{f})
=
\mathrm{d}[\ket{\psi(t_{f})},\ket{\psi_{\mathrm{aia}-1}(t_{f})}]
$,
using the transition times $\tau_{1,\pm}$, is plotted in
Fig.~\ref{fig:lzdaia} (red dotted line).
One can see that the AIA provides a slightly better approximation to
the fully evolved state than the adiabatic expansion only for very
small $t_{f}$.
For large $t_{f}$ we find,
$\mathrm{d}_{\mathrm{aia}-1}(t_{f})=99.08~t_{f}^{-1}$, providing a
much worst approximation than the adiabatic one, which is
$\mathrm{d}_{\mathrm{adi}}(t_{f})=0.074~t_{f}^{-1}$.
We attribute this discrepancy to the overestimation of the impulse
interval, given by the Kibble-Zurek argument~(\ref{eq:schroedeqlz}), in
the adiabatic limit ($t_{f}\to\infty$).

\subsubsection{Switching instants $\tau_{2,\pm}$: modified Kibble-Zurek argument}

Consequently, as a second scenario we consider a slight modification
of the Kibble-Zurek argument.
Namely, we replace the inverse rate of change of the external parameter
$z/\pd_{t}z$ in Eq.~(\ref{eq:kzatau}) by the inverse rate of change of the
Hamiltonian.
The resulting condition becomes
\be
\label{eq:kznatau}
\frac{1}{\Delta(\tau)}
=
\left.
\frac{\norm{\op{H}_{\mathrm{LZ}}(t)}_{\infty}}
     {\norm{\pd_{t}\op{H}_{\mathrm{LZ}}(t)}_{\infty}}
\right|_{t=\tau},
\ee
where $\norm{\cdot}_{\infty}$ is the operator norm defined by
$\norm{\op{O}}_{\infty} \equiv\max_{i} \mathrm{s}_{i}(\op{O})$, and
$\mathrm{s}_{i}(\op{O})$ are the singular values of $\op{O}$, i.e., eigenvalues
of $|\op{O}|\equiv\sqrt{\op{O}^{\dag}\op{O}}$.
The two real solutions of Eq.~(\ref{eq:kznatau}) are
\be
\tilde{\tau}_{2,\pm}
=
-\frac{z_{i}}{\delta\!z}t_{f}
\pm
\frac{x}{\sqrt{2} \, \delta\!z}
t_{f}
\sqrt{-2 + \frac{\delta\!z}{x^{2}}\frac{1}{t_{f}}},
\ee
and hence we obtain for the adiabatic-impulse switching times 
\be
\tau_{2,\pm}
=
\begin{cases}
  \begin{cases}
    t_{f}
    \\
    0
  \end{cases},
  &
  \qquad
  0 < t_{f} < \frac{1}{2}\frac{\delta\!z}{x^{2}+z_{i}^{2}}
  \\
  \tilde{\tau}_{2,\pm}, 
  &
  \qquad
  \frac{1}{2}\frac{\delta\!z}{x^{2}+z_{i}^{2}} < t_{f} < \frac{1}{2}\frac{\delta\!z}{x^{2}}
  \\
  \frac{t_{f}}{2},
  &
  \qquad
  \frac{1}{2} \frac{\delta\!z}{x^{2}} < t_{f} < \infty
\end{cases}.
\ee
The corresponding impulse interval reads,
$\Delta\tau_{2}\equiv\tau_{2,+}-\tau_{2,-}$,
and is also shown in Fig.~\ref{fig:lzdeltatau} (dashed green line).
It vanishes, if $t_{f}>\delta\!z/(2x^{2})$, and therefore
the distance
$
\mathrm{d}_{\mathrm{aia}-2}(t_{f})
=
\mathrm{d}[\ket{\psi(t_{f})}, \ket{\psi_{\mathrm{aia}-2}(t_{f})}]
$,
where the switching times $\tau_{2,\pm}$ are used,
recovers the adiabatic approximation at $t_{f}=\delta\!z/(2x^{2})$ (see
Fig.~\ref{fig:lzdaia} dashed green line).

\subsubsection{Switching instants $\tau_{3,\pm}$: simple gap condition}

Within the third approach, we consider the time, when the adiabatic
approximation fails, as the instant determining adiabatic-impulse
switching times $\tau_{\pm}$.
The simplest and crudest estimate for the adiabatic evolution to be
valid, might be given by $t_{f} \gg 1/\Delta$, as a consequence we
propose the simple equation
\be
\frac{1}{\Delta(\tau)}
=
t_{f},
\ee
to determine the time instants when the adiabaticity breaks down.
We find the two solutions
\be
\tilde{\tau}_{3,\pm}
=
-\frac{z_{i}}{\delta\!z} t_{f}
\pm
\frac{x}{\delta\!z} t_{f} \sqrt{-1 + \lp \frac{1}{2 x} \frac{1}{t_{f}}\rp^{2}},
\ee
from which we get for the adiabatic-impulse switching times
\be
\tau_{3,\pm}
=
\begin{cases}
  \begin{cases}
    t_{f}
    \\
    0
  \end{cases},
  &
  \qquad
  0 < t_{f} < \frac{1}{2} \frac{1}{\sqrt{x^{2}+z_{f}^{2}}}
  \\
  \tilde{\tau}_{3,\pm},
  &
  \qquad
  \frac{1}{2} \frac{1}{\sqrt{x^{2}+z_{f}^{2}}} < t_{f} < \frac{1}{2x}
  \\
  -\frac{z_{i}}{\delta\!z} t_{f},
  &
  \qquad
  \frac{1}{2x} < t_{f} < \infty
\end{cases}.
\ee
Figure~\ref{fig:lzdeltatau} shows also a plot of,
$\Delta\tau_{3}\equiv\tau_{3,+}-\tau_{3,-}$ (blue dot-dashed line).
In the present case the impulse interval vanishes for $t_{f}>1/(2x)$, 
which is much smaller than in the previous case, and hence the
resulting distance
$
\mathrm{d}_{\mathrm{aia}-3}(t_{f})
=
\mathrm{d}[\ket{\psi(t_{f})}, \ket{\psi_{\mathrm{aia}-3}(t_{f})}]
$,
becomes the same as for the adiabatic approximation at
$t_{f}=1/(2x)$ (see Fig.~\ref{fig:lzdaia} blue dot-dashed line).
This scenario does also not provide an improvement of the AIA, since
the estimate of the time when the adiabatic approximation fails is by
far underestimated.

\subsubsection{Switching instants $\tau_{4,\pm}$: ``folklore'' adiabatic condition}

\begin{figure}
\includegraphics[scale=0.9,trim=0.05mm 0.05mm 0.05mm 0.05mm, clip]{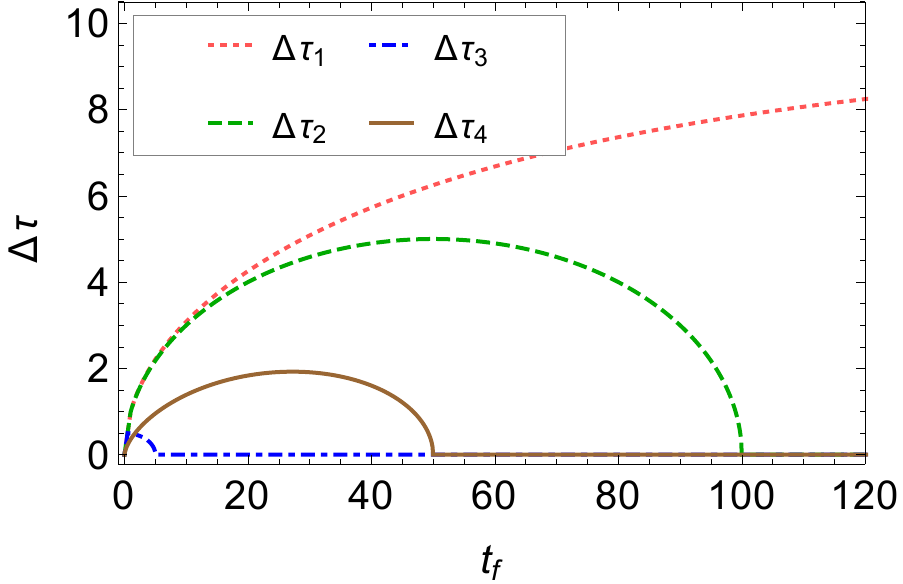}
\caption{
  (Color online)
  We show the impulse interval $\Delta\tau(t_{f})$ for $z_{i}=-1$,
  $z_{f}=1$ and $x=0.1$, found by the four different scenarios determining
  the impulse instants $\tau_{\pm}$.
  (1) The red dotted line shows $\Delta\tau_{1}$, obtained
  by the Kibble-Zurek argument $1/\Delta=z/\pd_{t}z$.
  (2) A modified Kibble-Zurek argument $1/\Delta=\norm{\op{H}}/\norm{\pd_{t}\op{H}}$ provided
  $\Delta\tau_{2}$, which is shown by the green dashed line.
  (3) Using the breakdown of the adiabatic theorem as an estimate for the
  impulse instants, $1/\Delta=t_{f}$, we found
  $\Delta\tau_{3}$, depicted by the blue dot-dashed.
  Finally, (4) the solid brown line shows $\Delta\tau_{4}$,
  given by the adiabaticity condition
  $\abs{\bra{\psi_{2}}\pd_{t}\op{H}\ket{\psi_{1}}}=\Delta^{2}$.  
  }
\label{fig:lzdeltatau}
\end{figure}

\begin{figure}
\includegraphics[scale=0.9,trim=0.05mm 0.05mm 0.05mm 0.05mm, clip]{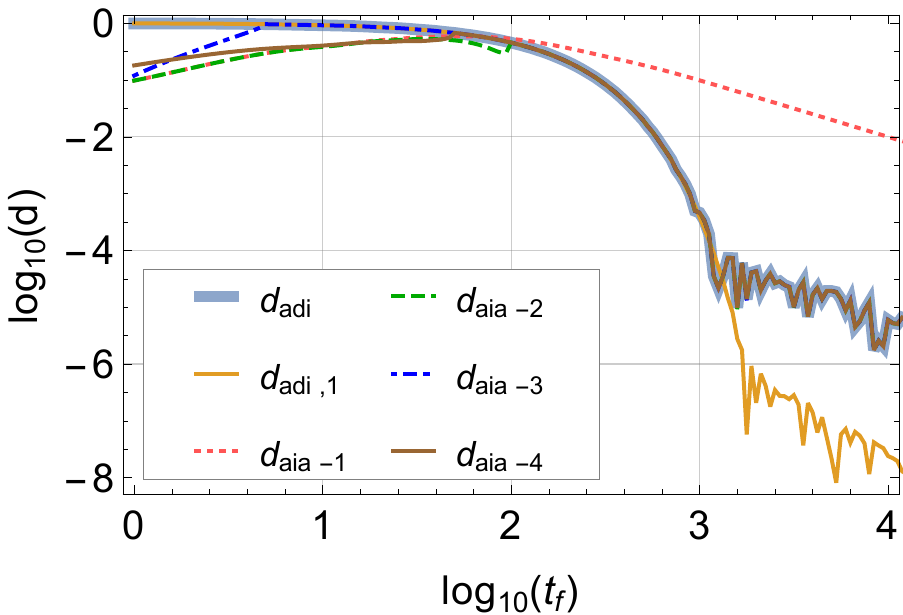}
\caption{
  (Color online)
  The distances between the AIA and the fully evolved state for the
  four different scenarios considered in the main text is plotted,
  with $z_{i}=-1$, $z_{f}=1$ and $x=0.1$.
  As a reference we also plotted the adiabatic distance
  $\mathrm{d}_{\mathrm{adi}}$ and the next first-order 
  correction $\mathrm{d}_{\mathrm{adi},1}$. 
  }
\label{fig:lzdaia}
\end{figure}

A more refined estimate for the validity of the adiabatic evolution, if
the system is initially prepared in the ground-state, might be provided by
\be
\max_{t \in \lb 0,t_{f} \rb}
\frac{\abs{\bra{\psi_{2}(t)} \pd_{t}\op{H}_{\mathrm{LZ}}(t) \ket{\psi_{1}(t)}}}
     {\Delta^{2}}
\ll 1,
\ee
see~\cite{Messiah1962}.
Consequently, we propose the following equation
\be
\label{eq:baalz}
\abs{\bra{\psi_{2}(t)} \pd_{t} \op{H}_{\mathrm{LZ}}(t) \ket{\psi_{1}(t)}}_{t=\tau} 
=
\Delta(\tau)^{2},
\ee
to determine the adiabatic-impulse switching times.
Equation~(\ref{eq:baalz}) has the two solutions
\be
\tilde{\tau}_{4,\pm}
=
-\frac{z_{i}}{\delta\!z} t_{f}
\pm
\frac{x}{\sqrt{2} \, \delta\!z} t_{f}
\sqrt{-2 + \lp \frac{\delta\!z}{\sqrt{2} \, x^{2}} \frac{1}{t_{f}}\rp^{2/3} },
\ee
which provides the adiabatic-impulse switching times
\be
\tau_{4,\pm}
=
\begin{cases}
  \begin{cases}
    t_{f}
    \\
    0
  \end{cases},
  &
  \qquad
  0 < t_{f} < \frac{1}{4}\frac{x \delta\!z}{x^{2}+z_{i}^{2}}
  \\
  \tilde{\tau}_{4,\pm},
  &
  \qquad
  \frac{1}{4}\frac{x \delta\!z}{x^{2}+z_{i}^{2}} < t_{f} < \frac{1}{4}\frac{\delta\!z}{x^{2}}
  \\
  \frac{t_{f}}{2},
  &
  \qquad
  \frac{1}{4}\frac{\delta\!z}{x^{2}} < t_{f} < \infty
\end{cases}.
\ee
Likewise, we plot the impulse interval,
$\Delta\tau_{4}\equiv\tau_{4,+}-\tau_{4,-}$,
in Fig~\ref{fig:lzdeltatau}, which is depicted by the brown solid line.
The interval now vanishes for $t_{f}>\delta\!z/(4 x^{2})$, which lies
in between the one found by the Kibble-Zurek argument using the
Hamiltonian's inverse rate of change and the interval found by the simple
adiabaticity breaking argument, $1/\Delta = t_{f}$.
As in the previous case, the resulting distance
$
\mathrm{d}_{\mathrm{aia}-4}(t_{f})
=
\mathrm{d}[\ket{\psi(t_{f})}, \ket{\psi_{\mathrm{aia}-4}(t_{f})}]
$,
recovers the adiabatic approximation, but now at
$t_{f}=\delta\!z/(4 x^{2})$.
This estimate of the impulse regime does therefore also not show any
major improvement of the AIA (solid brown line
in Fig.~\ref{fig:lzdaia}).
%
%
%

\subsubsection{Switching instants $\tau_{\mathrm{opt},\pm}$: optimization}

\begin{figure}
\includegraphics[scale=0.65,trim=0.01mm 0.01mm 0.01mm 0.01mm, clip]{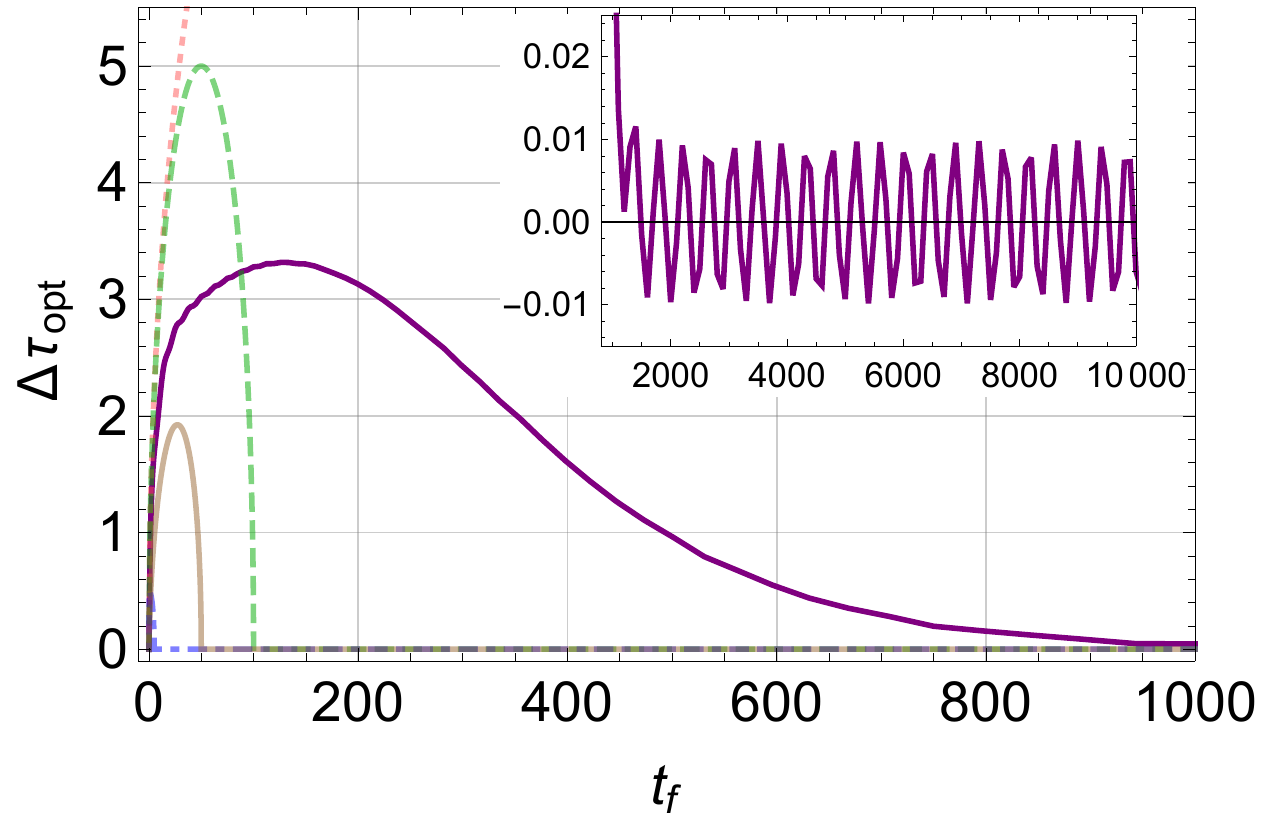}
\caption{
  (Color online)
  A plot of the optimal impulse interval,
  $\Delta\tau_{\mathrm{opt}}(t_{f})$, obtained by numerically
  minimizing $\mathrm{d}_{\mathrm{aia}}$, for $z_{i}=-1$,
  $z_{f}=1$ and $x=0.1$.
  For comparison we also show the impulse intervals found by the four
  different scenarios considered in the main text. 
  }
\label{fig:lzdtauopt}
\end{figure}

\begin{figure}
\includegraphics[scale=0.9,trim=0.05mm 0.05mm 0.05mm 0.05mm, clip]{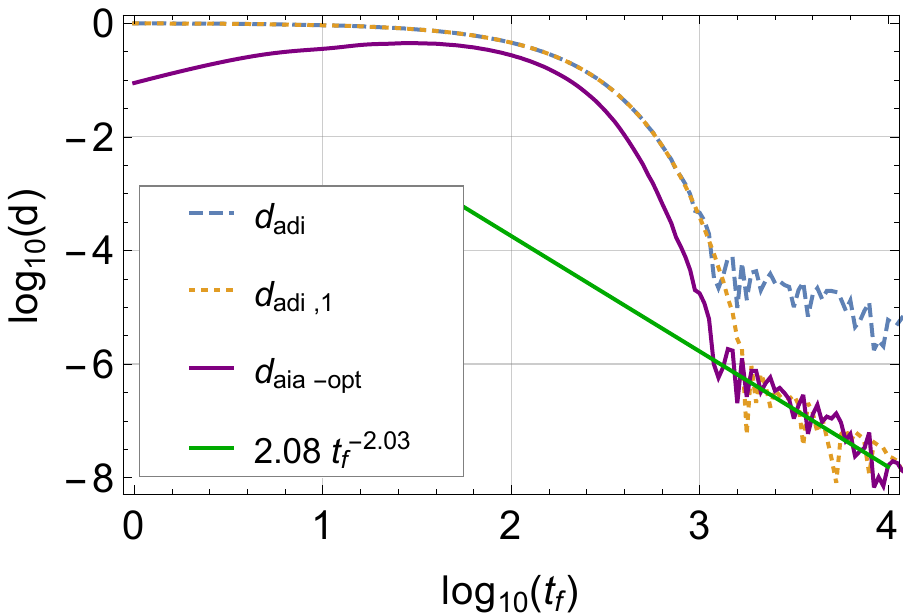}
\caption{
  (Color online)
  We show the minimal distance attainable by the adiabatic-impulse
  approximation, $\mathrm{d}_{\mathrm{aia-opt}}(t_{f})$, for $z_{i}=-1$,
  $z_{f}=1$ and $x=0.1$.
  As a reference we also plotted the adiabatic distance
  $\mathrm{d}_{\mathrm{adi}}$ and the first-order correction
  $\mathrm{d}_{\mathrm{adi},1}$.
  }
\label{fig:lzdaiaopt}
\end{figure}

Neither of the scenarios determining the impulse interval, which we
studied above, show an improvement with respect to the simple
adiabatic approximation.
One might therefore wonder, if there exist an optimal length of the
impulse interval, such that the AIA provides a better approximation to
the time evolved state than the adiabatic.
Consequently, we minimized the distance $\mathrm{d}_{\mathrm{aia}}$,
with respect to the impulse interval $\Delta\tau$, where we set 
$\tau_{\mathrm{opt},\pm} = t_{f}/2 \pm \Delta\tau/2$.
The numerically obtained result of $\Delta\tau_{\mathrm{opt}}(t_{f})$
is depicted in Fig.~\ref{fig:lzdtauopt}, for $z_{i}=-1$,
$z_{f}=1$ and $x=0.1$.
The functional form of $\Delta\tau_{\mathrm{opt}}(t_{f})$ is similar
to
$\pi \frac{1}{2}\frac{x^{2}}{\Delta z} t_{f} \exp{\lp-\pi
  \frac{1}{2}\frac{x^{2}}{\Delta z} t_{f}\rp}$,
although to our surprise, in the limit of large $t_{f}$ the optimal
interval manifests an oscillatory behavior around zero, which means
that it can become negative (Fig.~\ref{fig:lzdtauopt} inset).
This shows that after a certain final time it can become favorable to
pass the avoided level crossing adiabatically up to $\tau_{+}$, then make the
``impulse jump'' back to $\tau_{-}$, and finally go again through the avoided
level crossing adiabatically.
%
The resulting distance,
$
\mathrm{d}_{\mathrm{aia-opt}}(t_{f})
=
\mathrm{d}[\ket{\psi(t_{f})}, \ket{\psi_{\mathrm{aia-opt}}(t_{f})}]
$, 
is shown in Fig.~\ref{fig:lzdaiaopt} (solid purple line).
As a comparison, we also plotted the adiabatic distance
$\mathrm{d}_{\mathrm{adi}}$ and the first-order correction
$\mathrm{d}_{\mathrm{adi},1}$.
We can see an overall improvement of the AIA compared to the adiabatic
ones.
More surprisingly, we find that
$\mathrm{d}_{\mathrm{aia-opt}}(t_{f})=2.08~t_{f}^{-2.03}$, as
for the distance obtained by the first order adiabatic correction.
%

\subsection{Transverse Field Ising model}
\label{subsec:tfim}

In the following we will study the accuracy of the AIA for a closed
quantum many-body system.
We consider the illustrative example of the 
transverse field Ising (TFI)
model, whose Hamiltonian is given by
\be
\op{H}_{\mathrm{TFI}}(t)
=
\sum_{j=1}^{L} \op{\sigma}_{j}^{x}\op{\sigma}_{j+1}^{x}
+
h(t)
\sum_{j=1}^{L} \op{\sigma}_{j}^{z},
\ee
where $\op{\sigma}_{j}^{\alpha}$, with $\alpha=x,y,z$, are the Pauli
matrices describing the spin on the $j$-th site of the chain.
We assume periodic boundary conditions,
$\op{\sigma}_{L+1}^{\alpha}=\op{\sigma}_{1}^{\alpha}$,
such that the system remains translation invariant.
$h(t) \geq 0$ is the transverse magnetic field $h(t)$
acting in the $z$-direction and $L$ gives the total number of spins in
the chain.
For convenience we choose $L$ to be even.
We note that the Jordan-Wigner mapping separates the Hamiltonian into
two sub-spaces with an even or an odd number of fermions.
In the odd sector, the fermions satisfy periodic boundary conditions,
whereas in the even sector they obey anti-periodic boundary
conditions.
The Jordan-Wigner fermions are always created/destroyed in pairs, and
therefore the even/oddness of their number is conserved~\cite{Lieb1961}.
Consequently, we can fix a particular fermionic parity (here even),
which provides a unique ground-state $\ket{\Psi_{\mathrm{GS}}(h)}$.
In the even sector $\op{H}_{\mathrm{TFI}}$ can be mapped to a
non-interacting spin$-$1/2 model using a Jordan-Wigner followed by a
Fourier transformation: 
$\op{H}_{\mathrm{TFI}} = \sum_{k} {\bf c}_{k}^{\dag}
\tilde{H}_{k}^{\phantom{\dag}} {\bf c}_{k}^{\phantom{\dag}}$,
where
\be
\tilde{H}_{k}
=
-
\begin{pmatrix}
h - \cos k &  -i \, \sin k \\
i \, \sin k & - (h - \cos k)
\end{pmatrix},
\ee
with the pseudo-momenta given by
\be
k 
=
\pm \frac{1}{2} \frac{2 \pi}{L},
\pm \frac{2}{2} \frac{2 \pi}{L},
\ldots
\pm \lp \frac{L}{2}-\frac{1}{2} \rp \frac{2 \pi}{L},
\ee
and ${\bf c}_{k}^{\dag}=(\op{c}_{-k}^{\phantom{\dag}}, \op{c}_{k}^{\dag})$, with
$\op{c}_{k}$ being the Fourier transform of the Jordan-Wigner
fermions~\cite{Lieb1961, Henkel1999}.
As a consequence, the dynamics of the transverse field Ising model can
be decomposed into a collection of uncoupled two-level
systems~\cite{Dziarmaga2005, Cherng2006}.
Finally, through a Bogoliubov transformation, $\op{H}_{\mathrm{TFI}}$, can be
mapped to a free fermionic Hamiltonian,
$\op{H}_{\mathrm{TFI}}
=
\sum_{k}
\epsilon_{k}
(\op{\gamma}_{k}^{\dag}\op{\gamma}_{k}^{\phantom{\dag}}-\frac{1}{2})$,
with excitation spectrum
\be
\epsilon_{k} = 2 \sqrt{(h - \cos k)^{2} + \sin^{2}k},
\ee
and
$
\op{\gamma}_{k}
=
\cos\frac{\theta_{k}}{2}\op{c}_{k}^{\phantom{\dag}}-i\sin\frac{\theta_{k}}{2}\op{c}_{-k}^{\dag}
$,
where 
$
\theta_{k}
=
\arctan(\frac{\sin k}{h - \cos k})
$.
The ground-state of the transverse field Ising model is the vacuum of
the Bogoliubov operators, i.e., it is annihilated by all
$\op{\gamma}_{k}$, and thus
reads
\be
\ket{\Psi_{\mathrm{GS}}(h)}
=
\prod_{k}
\lp
\cos\frac{\theta_{k}}{2}\ket{0}_{k}\ket{0}_{-k}
+
i
\sin\frac{\theta_{k}}{2}\ket{1}_{k}\ket{1}_{-k}
\rp,
\ee
where
$\ket{1}_{k}=c_{k}^{\dag}\ket{0}_{k}$.
The corresponding ground-state energy is given by
$
E_{\mathrm{GS}}
=
-\frac{1}{2} \sum_{k} \epsilon_{k},
$
which in the thermodynamic limit ($L\to\infty$) becomes
\be
E_{\mathrm{GS}}
=
-\frac{L}{2\pi} \int_{0}^{\pi} dk \epsilon_{k}
=
-\frac{L}{2\pi} 2 (1+h) \mathcal{E}\lb\frac{4 h}{(1+h)^{2}}\rb,
\ee
where $\mathcal{E}\lb m \rb\equiv\int_{0}^{\pi/2}dx\sqrt{1-m \sin^{2}x}$ is
the complete elliptic integral.
The energy of a single excitation, i.e., a state of the form 
$\ket{\Psi_{q}}=\op{\gamma}_{q}^{\dag}\ket{\Psi_{\mathrm{GS}}}$, is
$E_{q}=\epsilon_{q}+E_{\mathrm{GS}}$, and therefore the gap reads
$\Delta=E_{k_{0}}-E_{\mathrm{GS}}=\epsilon_{k_{0}}$, where $k_{0}$ is
the minimal momentum, defined by the minimum of the excitation energy
$\pd_{q}\epsilon_{q}=0$.
In the thermodynamic limit we have, $\Delta=\abs{h-1}$, and
thus the gap vanishes at $h_{c}=1$, which marks the quantum critical
point, where the system undergoes a quantum phase transition from a
paramagnetic phase $(h>1)$ to a ferromagnetic phase $(h<1)$.

We will use the schedule, $h(t)=h_{i}+(h_{f}-h_{i})t/t_{f}$,
with $t \in \lb 0,t_{f} \rb$, and the system initially prepared in the
ground-state, $\ket{\Psi(t=0)}=\ket{\Psi_{\mathrm{GS}}(h_{i})}$. 
The starting value $h_{i}$, is chosen to be in the ferromagnetic
phase, i.e., $h_{i}<1$, and the final value, $h_{f}>1$, in the
paramagnetic phase, such that the quantum critical point is crossed at
$h=h_{c}=1$.

\begin{figure}
\includegraphics[scale=0.88,trim=0.05mm 0.05mm 0.05mm 0.05mm, clip]{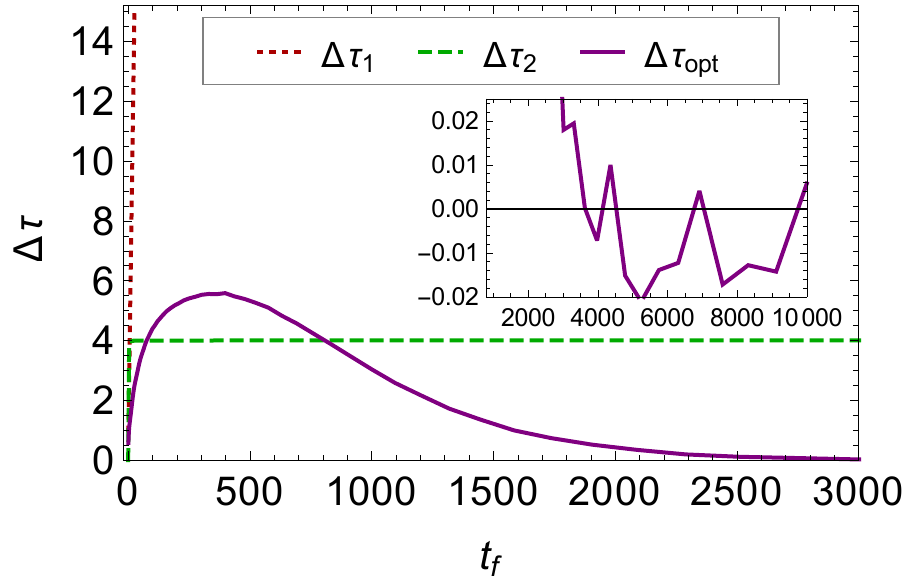}
\caption{
  (Color online)
  A plot of the impulse intervals found by the Kibble-Zurek argument,
  the Kibble-Zurek argument using the inverse rate of change of the
  Hamiltonian, and the optimal impulse interval found by minimizing
  $\mathrm{d}_{\mathrm{aia}}$ with respect to $\Delta\tau$.
  The initial field was $h_{i}=0.5$, the final $h_{f}=1.5$ and the
  minimization was performed for a chain of $L=150$.
  }
\label{fig:tfideltatau}
\end{figure}

In Fig.~\ref{fig:tfideltatau} we show the impulse interval
$\Delta\tau_{1}(t_{f})$, obtained by the Kibble-Zurek
argument $1/\Delta=h/\pd_{t}h$.
Solving this equation yields
\be
\tilde{\tau}_{1,\pm}
=
-\frac{h_{i}-1}{\delta h}t_{f}
\pm
\frac{1}{\sqrt{2}\sqrt{\delta h}} \sqrt{t_{f}}
\ee
with $\delta h \equiv h_{f}-h_{i}$, and the resulting impulse interval
reads 
\be
\Delta\tau_{1,\pm}
=
\begin{cases}
  t_{f},
  &
  \qquad
  0 < t_{f} < \frac{1}{2}\frac{\delta h}{(h_{f}-1)^{2}}
  \\
  \frac{\sqrt{2}}{\sqrt{\delta h}}\sqrt{t_{f}},
  &
  \qquad
  \frac{1}{2}\frac{\delta h}{(h_{f}-1)^{2}} < t_{f} < \infty
\end{cases}.
\ee
Further, we also plotted the impulse interval
$\Delta\tau_{2}(t_{t})$ in Fig.~\ref{fig:tfideltatau},
which is obtained by the modified Kibble-Zurek condition
\be
\frac{1}{\Delta}
=
\frac{\norm{\op{H}_{\mathrm{TFI}}}_{\infty}}{\norm{\pd_{t}\op{H}_{\mathrm{TFI}}}_{\infty}}.
\ee
Explicitly, we find
\be
\frac{1}{\abs{h(\tau)-1}}
=
\frac{[h(\tau)+1]\,\mathcal{E}\!\!\lp\frac{4h(\tau)}{[h(\tau)+1]^{2}}\rp}{\pi\,\pd_{t}h(t)|_{t=\tau}},
\ee
which we solved numerically to get $\Delta\tau_{2}(t_{t})$
(green dashed line in Fig.~\ref{fig:tfideltatau}).
Finally, we also plotted the impulse interval
$\Delta\tau_{\mathrm{opt}}(t_{t})$, found by minimizing the distance
between the fully evolved state and the AIA with respect to
$\Delta\tau$, 
where we set the impulse instants to
$\tau_{\mathrm{opt},\pm}=t_{f}/2\pm\Delta\tau/2$.
The result is depicted in Fig.~\ref{fig:tfideltatau} by a solid
purple.
Similar to the Landau-Zener model, we find that in the limit of large
$t_{f}$, the optimal impulse interval can become negative
(Fig.~\ref{fig:tfideltatau} inset).
Showing that for certain final times $t_{f}$, one can get a better
approximation to the fully evolved state by adiabatically crossing the
quantum criticality and evolve up to $\tau_{+}$, then make the
``impulse jump'' back to $\tau_{-}$, to finally go again through
the quantum phase transition.
%

\begin{figure}
\includegraphics[scale=0.9,trim=0.05mm 0.05mm 0.05mm 0.05mm, clip]{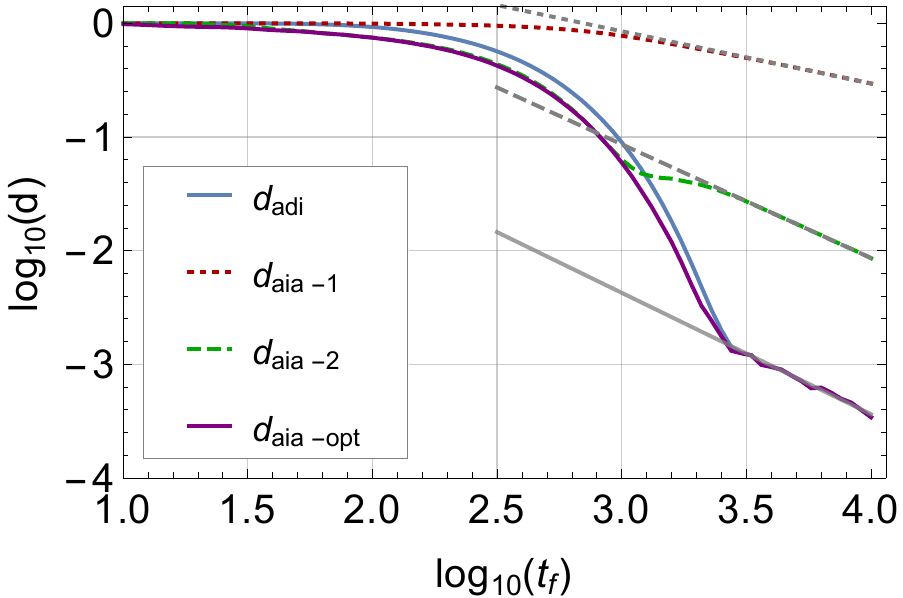}
\caption{
  (Color online)
  The distance between the numerically evolved ground-state and the
  different approximation schemes, i.e., adiabatic, adiabatic-impulse
  with the Kibble-Zurek argument, with the modified
  Kibble-Zurek argument, and the adiabatic-impulse approximation where
  the impulse instants are found by minimization of
  $\mathrm{d}_{\mathrm{aia}}$, is shown for $h_{i}=0.5$, $h_{f}=1.5$
  and $L=150$. 
  }
\label{fig:tfid}
\end{figure}

The corresponding distances $\mathrm{d}_{\mathrm{adi}}$,
$\mathrm{d}_{\mathrm{aia}-1}$, $\mathrm{d}_{\mathrm{aia}-2}$ and $\mathrm{d}_{\mathrm{aia-opt}}$
are plotted as a function of $t_{f}$ in Fig.~\ref{fig:tfid}, 
for the initial and final values $h_{i}=0.5$, $h_{f}=1.5$ and a chain
with $L=150$.
We observe the following large $t_{f}$ behavior,
$\mathrm{d}_{\mathrm{adi}}(t_{f}) = 6.8 \, t_{f}^{-1.07}$, 
$\mathrm{d}_{\mathrm{aia}-1}(t_{f}) = 20.3 \, t_{f}^{-0.46}$,
$\mathrm{d}_{\mathrm{aia}-2}(t_{f}) = 86.6 \, t_{f}^{-1.00}$ and
$\mathrm{d}_{\mathrm{aia-opt}}(t_{f}) = 6.8 \, t_{f}^{-1.07}$ (see
gray lines in Fig.~\ref{fig:tfid}).
The Kibble-Zurek argument gives an impulse interval that
grows with $\sqrt{t_{f}}$, and thus the corresponding distance,
$\mathrm{d}_{\mathrm{aia}-1}$, is always much larger than the
adiabatic one.
From this we conclude that $1/\Delta=\frac{h}{\pd_{t}h}$, clearly
overestimates the impulse interval.
However, we see
$\mathrm{d}_{\mathrm{aia}-2}<\mathrm{d}_{\mathrm{adi}}$, up to
$t_{f}=10^{3}$, showing that the modified Kibble-Zurek argument
yields a better estimate for the impulse interval. 
Although, for $t_{f} \geq 10^{3}$ we observe
$\mathrm{d}_{\mathrm{adi}}\ll\mathrm{d}_{\mathrm{aia}-2}$, which
implies that the impulse interval is still overestimated by the
modified Kibble-Zurek argument.
Obviously, the distance $\mathrm{d}_{\mathrm{aia-opt}}$, where the
impulse interval was found by minimizing the distance between the
AIA and the full evolution, gives the smallest distance.
Nevertheless, for our example of the transverse field Ising model the
improvement compared to the simple adiabatic approximation is
insignificant.
%

\section{Adiabatic-impulse approximation in open systems}
\label{sec:aiaopen}

In the following section we apply the AIA to the time evolution of
open quantum systems.
More specifically, we focus on dissipative systems characterized by a linear,
time-local master equation in the Lindblad form.
In a first step, we extend the AIA to the formalism used to describe
open quantum system.
To evaluate the accuracy of the AIA we will use the trace-norm
distance between the fully evolved density matrix and the approximated
one.
We use the adiabatic approximation as a reference to asses the
performance of the AIA.
As an example, we study the time evolution of a single qubit coupled to
a thermal bath, where the Liouvillian is in the Davies
form~\cite{Davies1974}.
The Davies generators arise in the limit of weak system-bath coupling.

\subsection{General setting}
\label{subsec:gsopen}

We consider an open quantum system of finite dimension, described by
the density matrix $\op{\rho}(t)$, whose evolution is governed by a
linear and time-local master equation
$\pd_{t} \op{\rho} = \widehat{\mathcal{L}}(t) \op{\rho}$.
The Liouvillian $\widehat{\mathcal{L}}(t)$ is in the Lindblad form
$
\widehat{\mathcal{L}}(t)\op{\rho}
=
-i\commutator{\op{H}(t)}{\op{\rho}}
+
\sum_{l}[\op{L}_{l}(t)\op{\rho}\op{L}_{l}^{\dag}(t)
       -
        \frac{1}{2}\anticommutator{\op{L}_{l}^{\dag}(t)\op{L}_{l}(t)}{\op{\rho}}]
$,
where $\op{H}(t)$ is the system Hamiltonian and $\{\op{L}_{l}(t)\}$ are the
Lindblad operators.
Further, we assume that the time dependence enters through the
parameter $\lambda(t)$.
The instantaneous steady states are defined by
$\widehat{\mathcal{L}}\op{\rho}_{1}=0$, 
and belong to the kernel of the Liouvillian.
We note that the Liouvillian operates on the space of linear
operators acting on the Hilbert space, which is denoted by
$\mathscr{L}(\mathcal{H})$.
This space can be turned into a Hilbert space,
when endowed with the Hilbert-Schmidt inner product
$\bbrackket{A}{B}=\mathrm{Tr}(\op{A}^{\dag}\op{B})$, for $\op{A},
\op{B} \in \mathscr{L}(\mathcal{H})$.
We notice, that for a properly normalized basis of hermitian
matrices $\{ \op{\Gamma}_{j} \}_{j=1}^{N}$, such that
$
\bbrackket{\Gamma_{i}}{\Gamma_{j}}
=
\mathrm{Tr}(\op{\Gamma}_{i}\op{\Gamma}_{j})
=
\delta_{ij}
$,
we can write the density matrix $\op{\rho}$ as $\kket{\rho} = \sum_{k=1}^{\dim\mathcal{H}^{2}}
c_{k} \kket{\Gamma_{k}}$, where $c_{k} = \bbrackket{\Gamma_{k}}{\rho} = \mathrm{Tr}(\op{\Gamma}_{k}\op{\rho})$.
The Liouvillian can therefore be interpreted as a matrix,
$\widehat{\mathcal{L}} = \sum_{j,k=1}^{\dim\mathcal{H}^{2}} \mathcal{L}_{jk} \kket{\Gamma_{j}}
\bbra{\Gamma_{k}}$, with the coefficients given by
$\mathcal{L}_{jk}
=
\bbra{\Gamma_{j}} \widehat{\mathcal{L}} \kket{\Gamma_{k}}
=
\mathrm{Tr}[\op{\Gamma}_{j}\widehat{\mathcal{L}}(\op{\Gamma}_{k})]$.
We assume that the Liouvillian has only semisimple eigenvalues, i.e.,
has no Jordan blocks, or in terms of the corresponding projectors 
$\widehat{\mathcal{L}}(t) \widehat{P}_{n}(t) = \lambda_{n}(t)
\widehat{P}_{n}(t)$.
This is guaranteed for the Davies generators, which we will consider
in the example below, since then $\widehat{\mathcal{L}}$ is normal.
The right and left eigenvectors of the Liouvillian are obtained by
\be
\widehat{\mathcal{L}} \kket{R_{n}^{(\alpha)}}
=
l_{n} \kket{R_{n}^{(\alpha)}},
\quad
\bbra{L_{m}^{(\alpha)}} \widehat{\mathcal{L}}
=
l_{m} \bbra{L_{m}^{(\alpha)}},
\ee
where $n,m=1,\ldots,\dim\mathcal{H}^{2}$ and $\alpha$ enumerates possible
degeneracies. 
The right eigenvector $\kket{R_{1}^{\alpha}}$ of the
eigenvalue $l_{1}=0$ are the instantaneous steady states in vector
notation.

We will consider the protocol, where the system at $t=0$ is
initialized in the instantaneous steady state $\op{\rho}_{1}(0)$,
and then we tune the parameter $\lambda(t)$, from $\lambda_{i}$ to
$\lambda(t_{f})=\lambda_{f}$, such that the gap of the Liouvillian
becomes minimal at a single instant in time.

Let us first recall the adiabatic approximation:
\be
\op{\rho}_{\mathrm{adi}}(t_{f})
=
\widehat{U}(t_{f},0) \op{\rho}_{1}(0),
\ee
where $\widehat{U}(t_{f},0)$ is the open system version of the full
adiabatic intertwiner, i.e., the operator that adiabatically all the
levels, see Appendix~\ref{appendix:B}.
Note that since $\op{\rho}_{1}(0) \in \ker\widehat{\mathcal{L}}$, we
have
$\widehat{U}(t_{f},0) \op{\rho}_{1}(0)
=
\widehat{W}_{1}(t_{f},0) \op{\rho}_{1}(0)$,
where $\widehat{W}_{1}(t_{f},0)$ evolves
adiabatically only vectors in the zero subspace (see
Appendix~\ref{appendix:A} and~\cite{CamposVenuti2016} for more details).

Let us now consider the AIA.
As in the closed case the evolution is assumed to be adiabatic from
$0$ to $\tau_{-}$, then it suddenly jumps from $\tau_{-}$ to
$\tau_{+}$
(in the region where the Liouvillian/Hamiltonian
gap is minimal), 
and finally becomes again adiabatic from $\tau_{+}$ to
$t_{f}$.
Consequently, the AIA can be written as
\be
\op{\rho}_{\mathrm{aia}}(t_{f})
=
\widehat{U}(t_{f},\tau_{+}) \, \op{1} \, \widehat{U}(\tau_{-},0) \op{\rho}_{1}(0).
\ee
At this point it is important to verify, whether the AIA map
$\widehat{U}(t_{f},\tau_{+}) \, \op{1} \, \widehat{U}(\tau_{-},0)$
is a bona fide completely positive trace preserving (CPTP) map.
In the Appendix~\ref{appendix:B} we show that indeed the full
intertwiner $\widehat{U}(t',t)$ is CPTP for $t' \geq t$, which in turn
implies that the AIA map is CPTP.
In doing so we actually prove an adiabatic theorem for the full
intertwiner $\widehat{U}$.
The whole complexity of the AIA lies in
the determination of the adiabatic-impulse switching times $\tau_{\pm}$.
For the Liouvillian in the Davies form we will simply use the
Hamiltonian gap as the relevant energy scale.

To measure the closeness of the AIA and
the adiabatic approximation to the time evolved state $\op{\rho}(t_{f})$,
we use the trace-norm distance, which is defined by
\be
\mathrm{d}(\op{\rho},\op{\sigma})
\equiv
\frac{1}{2} \norm{\op{\rho}-\op{\sigma}}_{1}
=
\frac{1}{2} \sum_{i} s_{i}(\op{\rho}-\op{\sigma}),
\ee
where $s_{i}(\op{X})$ are the singular values of $\op{X}$.
We note that for pure states the trace-norm distance reduces to the
distance~(\ref{eq:distclosed}).

\subsection{Single qubit coupled to a thermal bath}
\label{subsec:sqcttb}

We will study a single qubit coupled to a thermal bath at
inverse temperature $\beta=1/T$.
More specifically, the system Hamiltonian is assumed to be the
Landau-Zener model,
$\op{H}_{\mathrm{LZ}}(t) = x\op{\sigma}^{x} + z(t)\op{\sigma}^{z}$,
whose gap is given by $\Delta(t)=2\sqrt{x^{2}+z(t)^{2}}$.
The system-bath interaction is characterized by 
$\op{H}_{\mathrm{int}} = g \, \op{\sigma}^{y} \otimes \op{B}$, where $g$ is the
system-bath coupling constant, and $\op{B}$ some bath operator.
$\op{H}_{\mathrm{b}}$ describes the Hamiltonian of the bath.
Consequently, the total Hamiltonian reads
$\op{H}_{\mathrm{tot}}(t) = \op{H}_{\mathrm{LZ}}(t) +
\op{H}_{\mathrm{int}} + \op{H}_{\mathrm{b}}$.
We use a weak system-bath coupling and a slowly varying system
Hamiltonian~\cite{Albash2012}, therefore the time-dependent Lindblad
master-equation approximation describing the dynamics of the density
matrix $\op{\rho}(t)$, is assumed to be in the Davies
form~\cite{Davies1974},
\begin{align}
&
\widehat{\mathcal{L}}(t)\op{\rho}
=
-i \commutator{\op{H}_{\mathrm{LZ}}(t)}{\op{\rho}}
+
\nonumber \\
&
\sum_{\omega=\{0,\pm\Delta\}}
\gamma(\omega) 
[
\op{L}_{\omega}(t) \, \op{\rho} \, \op{L}_{\omega}^{\dag}(t)
-
\frac{1}{2}
\anticommutator{\op{L}_{\omega}^{\dag}(t) \, \op{L}_{\omega}(t)}{\op{\rho}}
],
\end{align}
where the spectral function of the bath $\gamma(\omega)$ is positive
and satisfies the Kubo-Martin-Schwinger (KMS) condition
$\gamma(-\omega)=e^{-\beta\omega}\gamma(\omega)$, see
Ref.~\cite{Kossakowski19771978}.
Let us choose $\gamma(\omega)$ to be in the Ohmic form
\be
\gamma(\omega)
=
2 \pi g^{2} \frac{\omega}{1-e^{-\beta \omega}}.
\ee
We note that the Davis form guarantees the steady states to be of
the Gibbs form,
$\op{\rho}_{1}=e^{-\beta\op{H}_{\mathrm{LZ}}}/Z$,
with $Z=\mathrm{Tr}_{S}(e^{-\beta \op{H}_{\mathrm{LZ}}})$,
see~\cite{Breuer2002} for more details.
The choice of $\op{H}_{\mathrm{int}} = g \, \op{\sigma}^{y} \otimes \op{B}$, 
ensure the minimum Lindbladian gap to be nonzero for all $z$, thus we
have as in the Landau-Zener case an avoided level crossing.
Further, we note that the Lamb shift Hamiltonian was neglected for
simplicity.
Finally, the Lindblad operators are given by
\be
\op{L}_{\omega}(t)
=
\sum_{i,j:~E_{i}-E_{j}=\omega}
\ket{\psi_{i}}\bra{\psi_{i}} \op{\sigma}^{y} \ket{\psi_{j}}\bra{\psi_{j}},
\ee
where $\omega \in \{ 0,\pm\Delta \}$, $i,j\in\{1,2\}$,
$E_{1,2}=\pm\sqrt{x^{2}+z^{2}}$ are the eigenenergies of the
Landau-Zener model and $\ket{\psi_{i}}$ denote the corresponding
eigenstates given in Eq.~(\ref{eq:eigvecslz}), and we obtain
\be
\op{L}_{0}
=0,
\quad
\op{L}_{+\Delta}
=
\frac{i\,z}{2\,b}\op{\sigma}^{x}
+
\frac{1}{2}\op{\sigma}^{y}
-
\frac{i\,x}{2\,b}\op{\sigma}^{z}
=
(\op{L}_{-\Delta})^{\dag}
.
\ee

The eigenvalues of the resulting Liouvillian
$\widehat{\mathcal{L}}(t)$ are given by 
\begin{align}
l_{1}
&=0,
\\
l_{2}
&=
-\lb\gamma(-\Delta)+\gamma(\Delta)\rb
=
- 2 \pi g^{2} \Delta \coth(\frac{\beta\Delta}{2}),
\\
l_{3}
&=
\frac{\lambda_{2}}{2} - i \Delta
=
-\pi g^{2} \Delta \coth(\frac{\beta\Delta}{2}) - i \Delta,
\\
l_{4}
&=
\frac{\lambda_{2}}{2} + i \Delta
=
-\pi g^{2} \Delta \coth(\frac{\beta\Delta}{2}) + i \Delta,
\end{align}
which are derived in Appendix~\ref{appendix:C}.
The corresponding left and right eigenvectors, denoted by
$\bbra{L_{i}}$ and $\kket{R_{i}}$,
respectively, with $i=1,2,3,4$, are also given in
Appendix~\ref{appendix:C}.

As for the closed case, we assume the protocol, where $x$ is constant
in time, $z(t)=z_{i}+(z_{f}-z_{i})\,t/t_{f}$, with
$t\in[0,t_{f}]$, and the system prepared in the state
$\op{\rho}(0)=\op{\rho}_{1}(0)$.
For the initial point, $z_{i}$, we choose a negative value, and for
the final point, $z_{f}$, the same but positive value, such that the
protocol passes the avoided level crossing (minimal gap) at $z=0$.
The time evolution is described by a linear, time-local master
equation of the form  $\pd_{t}\op{\rho}=\widehat{\mathcal{L}}(t)\op{\rho}$, 
which we express in the basis
$\frac{1}{\sqrt{2}}\{\op{1}, \op{\sigma}^{x}, \op{\sigma}^{y}, \op{\sigma}^{z}\}$
and solve numerically (see Appendix~\ref{appendix:D}).

First, we study the adiabatic approximation of the time evolved state, 
$\op{\rho}(t_{f})$, given by
\be
\op{\rho}_{\mathrm{adi}}(t_{f})
=
\widehat{U}(t_{f},0) \op{\rho}_{1}(0).
\ee
Note that since $\op{\rho}_{1}(0) \in \ker\widehat{\mathcal{L}}$ and
the latter is one dimensional, we have
\be
\kket{\rho_{\mathrm{adi}}(t_{f})}
=
\kket{R_{1}(t_{f})}.
\ee
The dynamical phase is zero, since $l_{1}=0$, and the Berry
phase is also zero, due to the fact that
$\bbra{L_{j}}\pd_{z}\kket{R_{j}}=0$, for $j=1,2,3,4$.

Within the AIA the time evolved state of our system is approximated by
\be
\op{\rho}_{\mathrm{aia}}(t_{f})
=
\widehat{U}(t_{f},\tau_{+}) \, \op{1} \, \widehat{U}(\tau_{-},0) \op{\rho}_{1}(0).
\ee
In vector notation we find
\be
\kket{\rho_{\mathrm{aia}}(t_{f})}
=
\sum_{j=1}^{4}
e^{\ell_{j}(\tau_{+},t_{f})}
\bbrackket{L_{j}(\tau_{+})}{R_{1}(\tau_{-})}
\kket{R_{j}(t_{f})},
\ee
where the dynamical phase reads,
$\ell_{j}(\tau_{+},t_{f})=\int_{\tau_{+}}^{t_{f}}dt \, l_{j}$,
and the Berry phases vanish as mentioned above.
To estimate the adiabatic-impulse switching times, $\tau_{\pm}$, we
will use the gap $\Delta$ of the system Hamiltonian, 
and therefore refer to the Sec.~\ref{subsec:lzm} for the estimation of
$\tau_{\pm}$.

\begin{figure}
\includegraphics[scale=0.5,trim=0.05mm 0.05mm 0.05mm 0.05mm, clip]{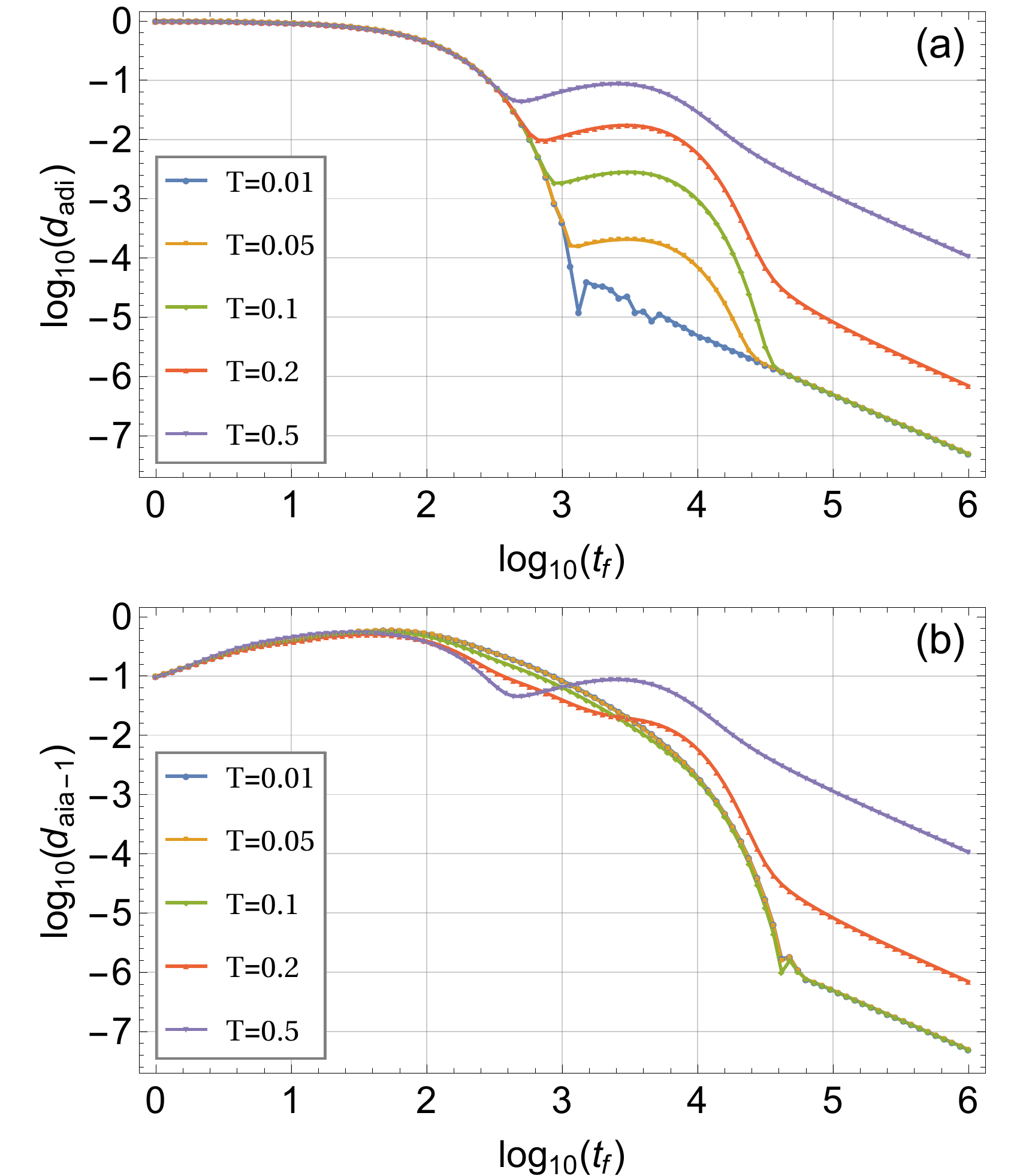}
\caption{
  (Color online)
  The trace-norm distance between the fully evolved state and the
  adiabatic/adiabatic-impulse approximation is shown.
  The panel (a) shows $\mathrm{d}_{\mathrm{adi}}(t_{f})$ and the panel
  (b) depicts $\mathrm{d}_{\mathrm{aia}-1}(t_{f})$ on a logarithmic
  scale.
  We plotted the trace-norm distance for different temperatures and
  for $x=0.1$, $z_{i}=-1$, $z_{f}=1$, and $g=0.01$. 
  }
\label{fig:scqttbdadiasa}
\end{figure}

The trace-norm distance between the exact evolution and the adiabatic
approximation/AIA are shown in Fig.~\ref{fig:scqttbdadiasa} for
different temperatures $T$, and for $x=0.1$, $z_{i}=-1$, $z_{f}=1$ and
$g=0.01$.
In Fig.~\ref{fig:scqttbdadiasa}~(a) we plot
$\mathrm{d}_{\mathrm{adi}}(t_{f})$ and in
Fig.~\ref{fig:scqttbdadiasa}~(b) $\mathrm{d}_{\mathrm{aia}-1}(t_{f})$,
where the impulse interval was estimated by the Kibble-Zurek argument
using the gap $\Delta$ of $\op{H}_{\mathrm{LZ}}$.
In contrast to the closed case, we observe that the trace-norm
distance $\mathrm{d}_{\mathrm{adi}}(t_{f})$ and
$\mathrm{d}_{\mathrm{aia}-1}(t_{f})$ become the same for large
$t_{f}$.
We believe an important ingredient to understand this phenomena is the
fact that the $\ker(\widehat{\mathcal{L}})$ is one-dimensional.

\begin{figure}
\includegraphics[scale=0.5,trim=0.05mm 0.05mm 0.05mm 0.05mm, clip]{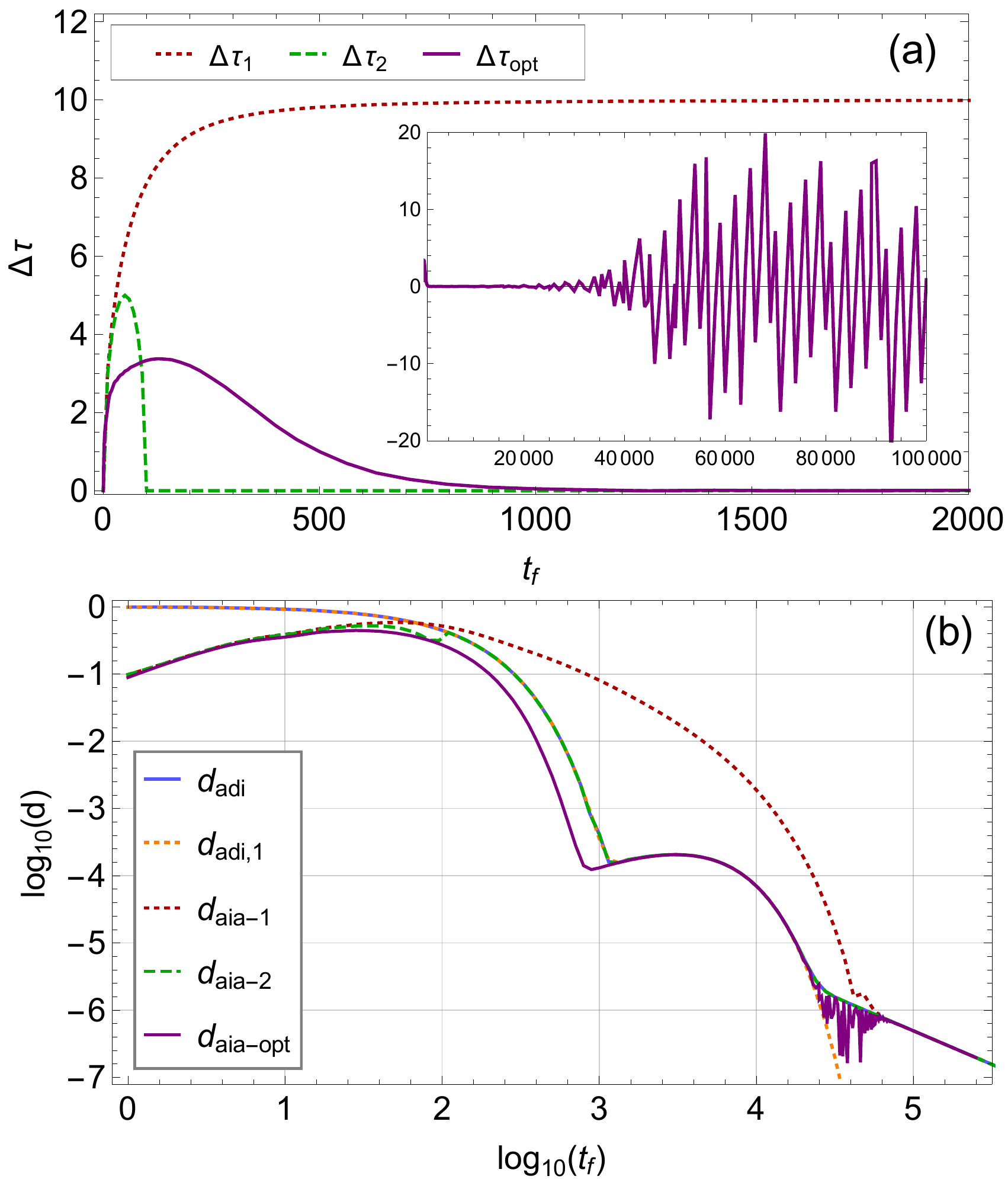}
\caption{
  (Color online)
  (a) We plot the impulse interval $\Delta\tau$ obtained by the
  Kibble-Zurek argument, the Kibble-Zurek argument using the rate of
  change of the Hamiltonian, and the optimal impulse interval found by
  minimizing the trace-norm distance $\mathrm{d}_{\mathrm{aia}}$ with
  respect to $\Delta\tau$. 
  (b) We compare the trace-norm distance found by the different
  scenarios mentioned in (a).
  The following values were used in both panels $T=0.05$, $x=0.1$,
  $z_{i}=-1$, $z_{f}=1$, and $g=0.01$.
  }
\label{fig:scqttbdeltatdadiasaopt}
\end{figure}

In Fig.~\ref{fig:scqttbdeltatdadiasaopt}~(a) we plot the impulse
interval $\Delta\tau_{1}$ obtained by the Kibble-Zurek
argument $1/\Delta=z/\pd_{t}z$, $\Delta\tau_{2}$ given by the modified
Kibble-Zurek argument 
$1/\Delta=\norm{\op{H}_{\mathrm{LZ}}}/\norm{\pd_{t}\op{H}_{\mathrm{LZ}}}$, 
and $\Delta\tau_{\mathrm{opt}}$ found by minimizing the trace-norm
distance $\mathrm{d}_{\mathrm{aia}}$ with respect to $\Delta\tau$.
The resulting trace-norm distances are compared in
Fig.~\ref{fig:scqttbdeltatdadiasaopt}~(b).
It is interesting to see that the for large $t_{f}$ all the
approximation schemes give the same trace norm distance as the
adiabatic approximation.
We observe although, that the first order correction adiabatic
correction provide still a smaller distance.
Even the trace norm distance found by the minimization process
becomes the same as the simple adiabatic approximation.
We note that the first order correction to the adiabatic approximation
can be found in Ref.~\cite{Avron2012} (Theorem 6).
However, there is a regime for which $\mathrm{d}_{\mathrm{opt}}$ can
reach the same distance as the first order adiabatic approximation, if
the counter intuitive scheme of crossing the minimal gap region twice
is applied.
%
%
%

\section{Conclusions}
\label{sec:conclusions}

We studied the accuracy of the AIA for closed and open quantum
systems, by evaluating how well this approximation reproduces the
exactly evolved state of the system. 
We used the trace-norm distance to characterize the closeness of the
approximated state to the exactly evolved one.
The adiabatic approximation served as a reference for the evaluation
of the AIA.
As might be expected, the AIA performs better than the adiabatic
approximation for small total evolution times $t_{f}$.
For large total evolution times, we observed that the Kibble-Zurek
argument overestimates the impulse interval and thus the AIA provided
a poor approximation to the time evolved state.

Modifying the Kibble-Zurek argument allowed us to improve the
AIA, such that at least the adiabatic approximation can be recovered.
However, the AIA can outperform the adiabatic one for large
$t_{f}$, if a counter intuitive procedure is applied.
Namely, driving the system adiabatically through the region where the
gap is minimal, then jumping back, to cross the minimal gap region
once again adiabatically.
We illustrated by several examples, that it is highly non trivial to
estimate the optimal impulse regime and even harder to guess, when to
cross the minimal gap twice, using this counter intuitive recipe.

We conclude, that the adiabatic-impulse approximation is a good method
to estimate the scaling behavior of certain non-equilibrium
properties, see, e.g.,~\cite{Damski2005, Damski2006, Dziarmaga2010}
for closed quantum systems or for dissipative quantum
systems~\cite{Nalbach2015, Patane2008, Keck2017}.
Nevertheless, to use it as a rigorous approximation for the time
evolution of quantum systems that are driven across a minimal gap
region, one still needs to get nontrivial knowledge about the system's
properties.

\section{Acknowledgments}

The research is based upon work (partially) supported by the Office of
the Director of National Intelligence (ODNI), Intelligence Advanced
Research Projects Activity (IARPA), via the U.S.\ Army Research Office
contract W911NF-17-C-0050.
The views and conclusions contained herein are those of the authors
and should not be interpreted as necessarily representing the official
policies or endorsements, either expressed or implied, of the ODNI,
IARPA, or the U.S.\ Government.
The U.S.\ Government is authorized to reproduce and distribute reprints
for Governmental purposes notwithstanding any copyright annotation
thereon.
This work was also supported by the Swiss National Science
Foundation (SNSF) and by the ARO MURI grant W911NF-11-1-0268.
M.T.\ is grateful to R.\ Di Felice and the CNR-NANO Institute in Modena,
Italy for their kind hospitality.
M.T.\ would also like to thank G.\ Styliaris and J.\ Marshall for helpful
discussions.
%


\begin{appendix}

\section{Adiabatic intertwiner for a single level}
\label{appendix:A}

In this appendix we express the adiabatic intertwiner
$\widehat{W}_{1}(t_{2},t_{1})$, which evolves adiabatically a single
instantaneous steady states of the Liouvillian from $t_{1}$ to
$t_{2}$, in vector notation.
In case the evolution is a long a closed loop the adiabatic
intertwiner becomes the so called Wilczek-Zee operator~\cite{Wilczek1984, Zanardi2015}.
The instantaneous steady states, i.e., the states from the kernel of
the Liouvillian $\widehat{\mathcal{L}}(t)$, are defined by
$\widehat{\mathcal{L}}(t)\op{\rho}_{1}^{(\alpha)}(t)=0$, where $\alpha$ enumerates
possible degeneracy.
In vector notation the instantaneous steady states
$\op{\rho}_{1}^{(\alpha)}(t)$ are the right eigenvectors of the
Liouvillian given by 
$\widehat{\mathcal{L}}(t)\kket{R_{1}^{(\alpha)}}=0$.
Further, the instantaneous spectral projection of
$\widehat{\mathcal{L}}(t)$ with zero eigenvalue is denoted by
$\widehat{P}_{1}(t)$.
We note that for a Liouvillian in the Lindblad form, the zero
eigenvalue (possibly degenerate) is semisimple, i.e., there are no
Jordan blocks associated to the zero eigenvalue and thus there are no
nilpotent terms in the zero sector
$\widehat{\mathcal{L}}\widehat{P}_{1}=\widehat{P}_{1}\widehat{\mathcal{L}}=0$, 
see~\cite{CamposVenuti2016} for a detailed proof.

The ideal adiabatic evolution is described by an operator
$\widehat{V}_{1}(t,0)$, satisfying the intertwining property
$\widehat{V}_{1}(t,0)\widehat{P}_{1}(0)=\widehat{P}_{1}(t)\widehat{V}_{1}(t,0)$
and is given by the solution of
\begin{align}
\pd_{t}\widehat{V}_{1}(t,0)
&=
\commutator{\pd_{t}\widehat{P}_{1}(t)}{\widehat{P}_{1}(t)}\widehat{V}_{1}(t,0)
\\
\widehat{V}_{1}(0,0)
&=
\op{1},
\end{align}
where $\op{1}$ is the identity operator.
One can see that $\widehat{V}_{1}(t,0)$ is not, in general, a completely
positive trace preserving (CPTP) map~\cite{CamposVenuti2016}, however
$\widehat{W}_{1}(t,0)\equiv\widehat{V}_{1}(t,0)\widehat{P}_{1}(0)$
is a CPTP map and thus the proper adiabatic intertwiner, e.g.,
$\op{\rho}_{1}^{(\alpha)}(t)=\widehat{W}_{1}(t,0)\op{\rho}_{1}^{(\alpha)}(0)$.
In addition, it was shown in~\cite{CamposVenuti2016}, that we can
write 
\be
\label{eq:w1prod}
\widehat{W}_{1}(t,0)
=
\lim_{N\to\infty}
\widehat{P}_{1}(N\epsilon)
\cdots
\widehat{P}_{1}(2\epsilon)
\widehat{P}_{1}(\epsilon)
\widehat{P}_{1}(0).
\ee
where ($t=N\epsilon$).
So we write
\be
\widehat{P}_{1}(t)
=
\sum_{\alpha}\kket{R_{1}^{(\alpha)}(t)}\bbra{L_{1}^{(\alpha)}(t)},
\ee
and note that
\begin{align}
&\bbrackket{L_{1}^{(\alpha_{j+1})}(t_{j+1})}{R_{1}^{(\alpha_{j})}(t_{j})}
\nonumber \\
&=
\delta_{\alpha_{j+1},\alpha_{j}}
+
\epsilon\,
\bbra{L_{1}^{(\alpha_{j+1})}(t)}
\overleftarrow{\partial}_{t}|_{t=t_{j}}\kket{R_{1}^{(\alpha_{j})}(t_{j})}
+
O\left(\epsilon^{2}\right)
\nonumber \\
&=
\delta_{\alpha_{j+1},\alpha_{j}}
-
\epsilon\,
\bbra{L_{1}^{(\alpha_{j+1})}(t_{j})}\partial_{t}\kket{R_{1}^{(\alpha_{j})}(t)}|_{t=t_{j}}
+
O\left(\epsilon^{2}\right)
\nonumber \\
&=
\left[\1-\epsilon A(t_{j})\right]_{\alpha_{j+1},\alpha_{j}}
+
O\left(\epsilon^{2}\right),
\label{eq:aexapnsion}
\end{align}
where $t_{j}=\epsilon\,j$, with $j = 0, 1, \ldots, N$ and $t_{N}=t$.
The second line follows by differentiating
$\bbrackket{L_{1}^{(\alpha)}}{R_{1}^{(\beta)}}=\delta_{\alpha,\beta}$
and we defined 
$[A_{1}(t)]_{\alpha,\beta}
\equiv
\bbra{L_{1}^{(\alpha)}(t)}\partial_{t}\kket{R_{1}^{(\beta)}(t)}$.
Plugging Eq.~(\ref{eq:aexapnsion}) into Eq.~(\ref{eq:w1prod}) yields
\begin{align}
\widehat{W}_{1}(t,0)
&=
\sum_{\alpha_{N},\alpha_{0}}
\kket{R_{1}^{(\alpha_{N})}(t)}\bbra{L_{1}^{(\alpha_{0})}(0)} 
\nonumber \\
&\phantom{=}
\,\left[
\overleftarrow{\mathrm{T}}\!\!\exp\lb-\int_{0}^{t} A_{1}(\sigma) d\sigma \rb
\right]_{\alpha_{N},\alpha_{0}},
\end{align}
where $\overleftarrow{\mathrm{T}}$ is the so called time-ordering
operator, ordering the operators in a chronological order with time
increasing from right to left.
Now we note further that, 
\be
\left\{\overleftarrow{\mathrm{T}}\!\!\exp\lb-\int_{0}^{t}A_{1}(\sigma)d\sigma\rb\right\}^{T}
=
\overrightarrow{\mathrm{T}}\!\!\exp\lb-\int_{0}^{t}A_{1}^{T}(\sigma)d\sigma\rb,
\ee
where $T$ indicates transpose and $\overrightarrow{\mathrm{T}}$
arranges operator in a chronological order, with time increasing from
left to right.
Finally, we can write
\begin{align}
\widehat{W}_{1}(t,0)
&=
\sum_{\alpha_{N},\alpha_{0}}
\kket{R_{1}^{(\alpha_{N})}(t)}\bbra{L_{1}^{(\alpha_{0})}(0)} \cdot
\nonumber \\
&\phantom{=}
\cdot
\,\left[
\overrightarrow{\mathrm{T}}\!\!\exp\lb-\int_{0}^{t}A^{T}(\sigma)d\sigma\rb
\right]_{\alpha_{0},\alpha_{N}},
\end{align}
which is the formula usually found in the literature~\cite{Wilczek1984, Zanardi2015}.

\section{The full adiabatic intertwiner}
\label{appendix:B}

In this appendix we want to show that the full adiabatic intertwiner $\widehat{U}$,
i.e., the map that evolves adiabatically all the levels and not only
a single one, is a bona fide completely positive trace preserving (CPTP)
map.
In doing so we will also prove the adiabatic theorem for $\widehat{U}$.
First, we note that it is convenient to rescale the time by the total
evolution time $t_{f}$, $s(t)=t/t_{f}$, such that $s\in[0,1]$.
Second, we remark that the dot will stand for differentiation with
respect to $s$, $\dot{X}=\pd_{s}X$.
Further, we assume the following spectral resolution of the
Liouvillian 
$\widehat{\mathcal{L}}(s)\widehat{P}_{n}(s)=l_{n}(s)\widehat{P}_{n}(s)$,
in other words we assume no Jordan blocks.
We also assume that all the levels $l_{n}(s)$ do not cross
and $\widehat{P}_{n}(s)$ are twice differentiable.

We begin by defining $V_{n}(s,s')$ as the solution of the following
ODE
\be
\dot{\widehat{V}}_{n}
=
\left(
t_{f}\widehat{\mathcal{L}}
+
[\dot{\widehat{P}}_{n},\widehat{P}_{n}]
\right)
\widehat{V}_{n},
~~
\widehat{V}_{n}(0)=\op{1}.
\ee
Differentiating
$
\widehat{h}(s,s')\equiv\widehat{V}_{n}(s,s')\widehat{P}_{n}(s')\widehat{V}_{n}(s',0)
$
with respect to $s'$, one sees that $\widehat{V}_{n}(s)$ has the
intertwining property:
\be
\widehat{P}_{n}(s)\widehat{V}_{n}(s)
=
\widehat{V}_{n}(s)\widehat{P}_{n}(0).
\ee
Let us define also
$\widehat{W}_{n}(s)\equiv\widehat{P}_{n}(s)\widehat{V}_{n}(s)$.
Using
$\dot{\widehat{P}}_{n}
=
\widehat{P}_{n}\dot{\widehat{P}}_{n}+\dot{\widehat{P}}_{n}\widehat{P}_{n}$,
one realizes that 
\begin{align}
\dot{\widehat{W}}_{n}
&=
(t_{f} \widehat{\mathcal{L}} + \dot{\widehat{P}}_{n}) \widehat{W}_{n}
\nonumber \\
&=\left( t_{f} \widehat{\mathcal{L}} + [\dot{\widehat{P}}_{n},\widehat{P}_{n}] \right) \widehat{W}_{n}.
\end{align}
Since $\widehat{W}_{n}$ satisfies the same ODE as $\widehat{V}_{n}$,
but with a different initial condition, 
we see that $\widehat{W}_{n}$ satisfies the intertwining property. 

Let us now further define
\be
\widehat{U}(s)\equiv\sum_{n}\widehat{W}_{n}(s).
\ee
We note that we assumed $\sum_{n}\widehat{P}_{n}=\op{1}$, i.e., that
the eigenvectors span the full space.
If this is not the case, i.e., there is also
a continuous spectrum, one can use the trick due to Kato, defining
the ``missing'' $\widehat{P}_{0}(s)$, such that the $\widehat{P}_{n}$
are then complete.
The differential equation for $\widehat{U}$ is
\begin{align}
\dot{\widehat{U}}
&=
\sum_{n}(t_{f}\widehat{\mathcal{L}}\widehat{P}_{n}
+
\dot{\widehat{P}}_{n}\widehat{P}_{n})\widehat{W}_{n}
\nonumber \\
&=
\sum_{n}(t_{f}\widehat{\mathcal{L}}\widehat{P}_{n}
+\dot{\widehat{P}}_{n}\widehat{P}_{n})\sum_{l}\widehat{W}_{l}.
\end{align}
Now using
$\sum_{n}\dot{\widehat{P}}_{n}\widehat{P}_{n}
=
-\sum_{n}\widehat{P}_{n}\dot{\widehat{P}}_{n}$,
which stems from the completeness of the $\widehat{P}_{n}$, one gets
\begin{align}
\label{eq:diffequ}
\dot{\widehat{U}}
&=
\left(t_{f}\widehat{\mathcal{L}}+\frac{1}{2}\sum_{n}[\dot{\widehat{P}}_{n},\widehat{P}_{n}]\right)\widehat{U},
\nonumber \\
\widehat{U}(0) & =\sum_{n}\widehat{P}_{n}(0)=\op{1}.
\end{align}
Clearly, $\widehat{U}(s)$ behaves like $\widehat{W}_{n}(s)$ in the
range of $\widehat{P}_{n}(0)$
for all $n$, so $\widehat{U}(s)$ might as well be called the full
intertwiner.

Let us now show that each $\widehat{V}_{n}$ is close to
$\widehat{\mathcal{E}}$ in the range of $\widehat{P}_{n}$,
where the operator $\widehat{\mathcal{E}}(s,0)$ is the evolution
operator, describing the full time evolution of the density matrix
$\op{\rho}(s)=\widehat{\mathcal{E}}(s,0)\op{\rho}(s)$,
and satisfies 
$\pd_{s}\widehat{\mathcal{E}}(s,0)=t_{f}\widehat{\mathcal{L}}(s)\widehat{\mathcal{E}}(s,0)$,
with $\widehat{\mathcal{E}}(s,s)=\op{1}$.
One has 
\begin{align}
&\widehat{\mathcal{E}}(0,s)\widehat{W}_{n}(s)-\widehat{P}_{n}(0)
=
\int_{0}^{s}ds'
\frac{d}{ds'}\left[\widehat{\mathcal{E}}(0,s')\widehat{W}_{n}(s')\right]
\nonumber \\
&=
\int_{0}^{s}ds'
\widehat{\mathcal{E}}(0,s')\dot{\widehat{P}}_{n}(s')\widehat{W}_{n}(s')
\nonumber \\
&=
\int_{0}^{s}ds'
\widehat{\mathcal{E}}(0,s')\widehat{Q}_{n}(s')\dot{\widehat{P}}_{n}(s')\widehat{W}_{n}(s'),
\end{align}
where $\widehat{Q}_{n}(s)=\op{1}-\widehat{P}_{n}(s)$, and by using the
identity $\widehat{P}_{n}\dot{\widehat{P}}_{n}\widehat{P}_{n}=0$.
The reduced resolvent is defined by
$\widehat{S}_{n}
=\lim_{a \to l_{n}}
\widehat{Q}_{n}
(\widehat{\mathcal{L}}-a\op{1})^{-1}
\widehat{Q}_{n}$,
which satisfies
$\widehat{Q}_{n}=\widehat{\mathcal{L}}\widehat{S}_{n}$,
and together with
\be
\widehat{\mathcal{E}}(0,s')
\widehat{\mathcal{L}}(s')
=-t_{f}^{-1}\partial_{s'}\widehat{\mathcal{E}}(0,s'),
\ee
implies
\begin{align}
&\widehat{\mathcal{E}}(0,s)\widehat{W}_{n}(s)-\widehat{P}_{n}(0)
\nonumber \\
&=
-
\frac{1}{t_{f}}
\int_{0}^{s}ds'
\left[\partial_{s'}\widehat{\mathcal{E}}(0,s')\right]
\widehat{S}_{n}(s')
\dot{\widehat{P}}_{n}(s')
\widehat{W}_{n}(s')
\nonumber \\
&=
-\frac{1}{t_{f}}
\left.
\widehat{\mathcal{E}}(0,s')
\widehat{S}_{n}(s')
\dot{\widehat{P}}_{n}(s')
\widehat{W}_{n}(s')
\right|_{0}^{s}
\nonumber \\
&+
\frac{1}{t_{f}}
\int_{0}^{s}ds'
\widehat{\mathcal{E}}(0,s')
\partial_{s'}\left[\widehat{S}_{n}(s')\dot{\widehat{P}}_{n}(s')\widehat{W}_{n}(s')\right].
\end{align}
We now multiply the last equation by $\widehat{\mathcal{E}}(s,0)$ from
the left and get
\begin{align}
&\widehat{\mathcal{E}}(s,0)\widehat{P}_{n}(0)
-
\widehat{V}_{n}(s)\widehat{P}_{n}(0)
\nonumber \\
&=
\frac{1}{t_{f}}
\left[
 \widehat{S}_{n}(s)
 \dot{\widehat{P}}_{n}(s)
 \widehat{W}_{n}(s)
 -
 \widehat{\mathcal{E}}(s,0)
 \widehat{S}_{n}(0)
 \dot{\widehat{P}}_{n}(0)
 \widehat{W}_{n}(0)
\right]
\nonumber \\
&-
\frac{1}{t_{f}}
\int_{0}^{s}ds'
\widehat{\mathcal{E}}(s,s')
\partial_{s'}\left[\widehat{S}_{n}(s')\dot{\widehat{P}}_{n}(s')\widehat{W}_{n}(s')\right].
\end{align}
We note that $\widehat{S}_{n}$ is the reduced resolvent of
$\widehat{\mathcal{L}}$ and does not contain $t_{f}$, so neither
$\widehat{S}_{n}$ nor $\widehat{P}_{n}$ depend on $t_{f}$.
However, in our formulation $\widehat{W}_{n}$ does depend on $t_{f}$. 
In~\cite{salem_quasi-static_2007} Salem simply claims
that $\widehat{W}_{n}$ is bounded.
This seems to overlook the fact that the constant for the bound could
still depend on $t_{f}$.
In any case, the required bound can be obtained by writing a Trotter
expansion for $\widehat{W}_{n}$:
\be
\widehat{W}_{n}(s)
=
\lim_{N\to\infty}
\overleftarrow{\mathrm{T}}
\prod_{i=1}^{N}
\left(
 e^{\epsilon t_{f} \widehat{\mathcal{L}}(s_{i})}
 e^{\epsilon \dot{\widehat{P}}_{n}(s_{i})}
\right)
\ee
with $\epsilon=s/N$, $s_{i}=\epsilon i$,
$\widehat{B}(s)\equiv\dot{\widehat{P}}_{n}(s)$.
This shows that $\norm{\widehat{W}_{n}}$ can be bounded by
a constant independent of $t_{f}$, since each $\widehat{\mathcal{L}}$
is a generator of a contraction semigroup.
In fact one obtains
\be
\left\Vert \widehat{W}_{n}(s) \right\Vert
\le
\exp\left(\int_{0}^{s}ds'\left\Vert \dot{\widehat{P}}_{n}(s') \right\Vert \right),
\ee
which shows finally that 
\be
\left\Vert 
 \left(\widehat{\mathcal{E}}(s)-\widehat{W}_{n}(s)\right)
 \widehat{P}_{n}(0)
\right\Vert
\le
\frac{C_{n}}{t_{f}},
\ee
where $C_{n}$ are finite constants independent of $t_{f}$.
Coming back to $\widehat{U}$, we can write
\begin{align}
\widehat{\mathcal{E}}(s) - \widehat{U}(s)
&=
\sum_{n}\lb \widehat{\mathcal{E}}(s) - \widehat{U}(s) \rb
\widehat{P}_{n}(0)
\nonumber \\
&=
\sum_{n}\lb\widehat{\mathcal{E}}(s)-\widehat{W}_{n}(s)\rb \widehat{P}_{n}(0),
\end{align}
and taking norms one obtains 
\be
\left\Vert \widehat{\mathcal{E}}(s) - \widehat{U}(s) \right\Vert
\le
\frac{1}{t_{f}} \sum_{n}C_{n}.
\ee
In finite dimension the latter sum $\sum_{n}C_{n}$ does not pose a
problem, since it is still finite.
Nevertheless, for infinite dimensional systems one should show that
the sum is bounded.
In summary, this implies that $\widehat{U}$ is arbitrarily close to a
CPTP map, and therefore is itself a CPTP map.
%

\section{Derivation of the eigenvalues and eigenvectors of the Liouvillian}

\label{appendix:C}

Let us now write the Liouvillian operator $\widehat{\mathcal{L}}$ in the
basis
$\{\op{\Gamma}_{i}\}_{i=1}^{4}
=
\frac{1}{\sqrt{2}}\{\op{1},\op{\sigma}^{x},\op{\sigma}^{y},\op{\sigma}^{z}\}$,
i.e.,
$\mathcal{L}_{ij}=\mathrm{Tr}(\op{\Gamma}_{i}\,\widehat{\mathcal{L}}\,\op{\Gamma}_{j})$,
we find
\begin{widetext}
\be
(\mathcal{L}_{ij})
=
\begin{pmatrix}
0
&
0
&
0
&
0
\\
\dfrac{2x}{\Delta} \lb\gamma(-\Delta)-\gamma(\Delta)\rb 
& 
-
\dfrac{2(x^{2}+\frac{1}{4}\Delta^{2})\lb\gamma(-\Delta)+\gamma(\Delta)\rb}{\Delta^{2}} 
&
-2 z
&
-\dfrac{2 x z \lb\gamma(-\Delta)+\gamma(\Delta)\rb}{\Delta^{2}}
\\
0
&
2z
&
- \dfrac{1}{2}\lb\gamma(-\Delta)-\gamma(\Delta)\rb
&
-2x
\\
\dfrac{2z\lb\gamma(-\Delta)-\gamma(\Delta)\rb}{\Delta}
&
- \dfrac{2 x z \lb\gamma(-\Delta)+\gamma(\Delta)\rb}{\Delta^{2}}
&
2 x
&
-\dfrac{2(\frac{1}{4}\Delta^{2}+z^{2})\lb\gamma(-\Delta)+\gamma(\Delta)\rb}{\Delta^{2}}
\end{pmatrix},
\ee
\end{widetext}
and the vector representation of the density matrix $\op{\rho}$ reads
\be
\kket{\rho}
=
\sum_{i=1}^{4} c_{i} \kket{\Gamma_{i}}
=
\begin{pmatrix}
c_{1} \\
c_{2} \\
c_{3} \\
c_{4}
\end{pmatrix},
\ee
where $c_{i}=\mathrm{Tr}(\op{\Gamma_{i}}\op{\rho})$.
The eigenvalues of the Liouvillian $\mathcal{L}$ can be calculated and
are given by
\begin{align}
l_{1}
&=0,
\\
l_{2}
&=
-\lb\gamma(-\Delta)+\gamma(\Delta)\rb
=
- 2 \pi g^{2} \Delta \coth(\frac{\Delta}{2}\beta),
\\
l_{3}
&=
-\frac{1}{2}
\lb\gamma(-\Delta)+\gamma(\Delta)\rb-i\Delta
\nonumber \\
&=
-\pi g^{2} \Delta \coth(\frac{\Delta}{2}\beta) - i \Delta,
\\
l_{4}
&=
-\frac{1}{2}
\lb\gamma(-\Delta)+\gamma(\Delta)\rb+i\Delta
\nonumber \\
&=
\pi g^{2} \Delta \coth(\frac{\Delta}{2}\beta) + i \Delta,
\end{align}
and the corresponding eigenvectors read
\begin{align}
\kket{r_{1}}
&=
\begin{pmatrix}
\frac{\Delta}{\sqrt{2}\,z} \frac{\gamma(-\Delta)+\gamma(\Delta)}{\gamma(-\Delta)-\gamma(\Delta)} \\
\frac{x}{z} \\
0 \\
1
\end{pmatrix}
=
\begin{pmatrix}
-\frac{\Delta}{\sqrt{2}\,z} \coth(\frac{\beta\Delta}{2}) \\
\frac{x}{z} \\
0 \\
1
\end{pmatrix},
\nonumber \\
\kket{r_{2}}
&=
\begin{pmatrix}
0 \\
\frac{x}{z} \\
0 \\
1
\end{pmatrix},
\quad
\kket{r_{3}}
=
\begin{pmatrix}
0 \\
-\frac{z}{x} \\
-i \frac{\Delta}{2\,x} \\
1
\end{pmatrix},
\quad
\kket{r_{4}}
=
\begin{pmatrix}
0 \\
-\frac{z}{x} \\
i \frac{\Delta}{2\,x} \\
1
\end{pmatrix}.
\end{align}
We note that they have to be normalized such that the corresponding
density matrices have trace one, since 
$\op{\rho}=\sum_{i}c_{i}\op{\Gamma}_{i}=c_{1}\frac{1}{\sqrt{2}}\op{1}+c_{2}\frac{1}{\sqrt{2}}\op{\sigma}^{x}+c_{3}\frac{1}{\sqrt{2}}\op{\sigma}^{y}+c_{4}\frac{1}{\sqrt{2}}\op{\sigma}^{z}$,
only the first component needs to be $\frac{1}{\sqrt{2}}$ in order to
have $\mathrm{Tr}(\op{\rho})=1$, because $\mathrm{Tr}(\op{1})=2$ and
$\mathrm{Tr}(\op{\sigma}^{\alpha})=0$ for $\alpha=x,y,z$. 
As we see this is only possible for the right eigenvector
$\kket{r_{1}}$ with $l_{1}=0$, and therefore we find
\begin{align}
\kket{\rho_{1}}
&=
\begin{pmatrix}
\frac{1}{\sqrt{2}} \\
-\frac{\sqrt{2}x}{\Delta} \tanh(\frac{\beta\Delta}{2}) \\
0 \\
-\frac{\sqrt{2}z}{\Delta} \tanh(\frac{\beta\Delta}{2})
\end{pmatrix},
\qquad
\kket{\rho_{2}}
=
\begin{pmatrix}
0 \\
\frac{x}{z} \\
0 \\
1
\end{pmatrix},
\nonumber \\
\kket{\rho_{3}}
&=
\begin{pmatrix}
0 \\
-\frac{z}{x} \\
-i \frac{\Delta}{2\,x} \\
1
\end{pmatrix},
\qquad
\kket{\rho_{4}}
=
\begin{pmatrix}
0 \\
-\frac{z}{x} \\
i \frac{\Delta}{2\,x} \\
1
\end{pmatrix}.
\end{align}
In matrix notation they read
\begin{align}
\op{\rho}_{1}
&=
\begin{pmatrix}
\frac{1}{2} - \frac{z \tanh(\frac{\beta\Delta}{2})}{\Delta}
&
- \frac{x \tanh(\frac{\beta\Delta}{2})}{\Delta}
\\
- \frac{x \tanh(\frac{\beta\Delta}{2})}{\Delta}
&
\frac{1}{2} + \frac{z \tanh(\frac{\beta\Delta}{2})}{\Delta}
\end{pmatrix}
=\frac{e^{-\beta \op{H}_{\mathrm{LZ}}}}{\mathrm{Tr}(e^{-\beta \op{H}_{\mathrm{LZ}}})},
\\
\op{\rho}_{2}
&=
\frac{1}{\sqrt{2}}
\begin{pmatrix}
1
&
\frac{x}{z}
\\
\frac{x}{z}
&
-1
\end{pmatrix},
\\
\op{\rho}_{3}
&=
\frac{1}{\sqrt{2}}
\begin{pmatrix}
1
&
- \frac{z}{x} - \frac{1}{2} \frac{\Delta}{x}
\\
- \frac{z}{x} + \frac{1}{2} \frac{\Delta}{x}
&
-1
\end{pmatrix},
\\
\op{\rho}_{4}
&=
\frac{1}{\sqrt{2}}
\begin{pmatrix}
1
&
- \frac{z}{x} + \frac{1}{2} \frac{\Delta}{x}
\\
- \frac{z}{x} - \frac{1}{2} \frac{\Delta}{x}
&
-1
\end{pmatrix}.
\end{align}
It can be seen that $\mathrm{Tr}(\op{\rho}_{2,3,4}) \neq 1$, and hence
$\op{\rho}_{2,3,4}$ can not be interpreted as states.
The Liouvillian gap is given by
\be
\Delta_{\mathrm{L}} = \mathrm{min} \{ \abs{l_{2}}, \abs{l_{3}} \},
\ee
where
\begin{align}
\abs{l_{2}} 
&=
\sqrt{\lambda_{2}^{\phantom{\ast}} \lambda_{2}^{\ast}} 
=
\sqrt{\lb\gamma(-\Delta)+\gamma(\Delta)\rb^{2}}
=
2 \pi g^{2} \Delta \coth(\frac{\beta\Delta}{2})
\nonumber \\
&=
4 \pi g^{2} T
+
\frac{1}{3} \pi g^{2} \frac{1}{T} \Delta^{2} +
\mathcal{O}(\Delta^{4})
\\
\abs{l_{3}} 
&=
\sqrt{\lambda_{3}^{\phantom{\ast}} \lambda_{3}^{\ast}} 
=
\sqrt{\frac{1}{4}\lb\gamma(-\Delta)+\gamma(\Delta)\rb^{2}+\Delta^{2}}
\nonumber \\
&=
2 \pi g^{2} T
+
\frac{3+2\pi^{2}g^{4}}{12 \pi g^{2}} \frac{1}{T} \Delta^{2}
+
\mathcal{O}(\Delta^{3}).
\end{align}
The left eigenvectors of the Liouvillian are defined by
\be
\bbra{l_{m}}\mathcal{L} = \lambda_{m} \bbra{l_{m}}
\iff
\mathcal{L}^{\dag} \kket{l_{m}} = l_{m}^{\ast} \kket{l_{m}},
\ee
and therefore we find for the left eigenvectors
\begin{align}
\bbra{l_{1}}
&=
\begin{pmatrix}
1, &
0, &
0, &
0
\end{pmatrix},
\\\
\bbra{l_{2}}
&=
\begin{pmatrix}
-\frac{\Delta}{2 \, z} \frac{\gamma(-\Delta)-\gamma(\Delta)}{\gamma(-\Delta)+\gamma(\Delta)}, &
\frac{x}{z}, &
0, &
1
\end{pmatrix}
\nonumber \\
&=
\begin{pmatrix}
\frac{\Delta}{2\,z} \tanh(\frac{\beta\Delta}{2}), &
\frac{x}{z}, &
0, &
1
\end{pmatrix},
\\
\bbra{l_{3}}
&=
\begin{pmatrix}
0, &
-\frac{z}{x}, &
i \frac{\Delta}{2\,x}, &
1
\end{pmatrix},
\\
\bbra{l_{4}}
&=
\begin{pmatrix}
0, &
-\frac{z}{x}, &
-i \frac{\Delta}{2\,x}, &
1
\end{pmatrix}.
\end{align}
We can normalize the first left eigenvector such that in matrix
notation we have
$
\bbrackket{\varpi_{1}}{\rho}
=
\mathrm{Tr}(\op{1}\op{\rho})=\mathrm{Tr}(\op{\rho})=1
$, i.e.,  
$
\bbra{\varpi_{1}}
=
\begin{pmatrix}
\sqrt{2}, &
0, &
0, &
0
\end{pmatrix}
$,
and thus we get
\begin{align}
\op{\varpi_{1}}
&=
\begin{pmatrix}
1
&
0
\\
0
&
1
\end{pmatrix}
\\
\op{\varpi}_{2}
&=
\frac{1}{\sqrt{2}}
\begin{pmatrix}
1 + \frac{\Delta}{2\,z} \tanh(\frac{\beta\Delta}{2})
&
\frac{x}{z}
\\
\frac{x}{z}
&
-1 + \frac{\Delta}{2\,z} \tanh(\frac{\beta\Delta}{2})
\end{pmatrix},
\\
\op{\varpi}_{3}
&=
\frac{1}{\sqrt{2}}
\begin{pmatrix}
1
&
- \frac{z}{x} + \frac{1}{2} \frac{\Delta}{x}
\\
- \frac{x}{z} - \frac{1}{2} \frac{\Delta}{x}
&
-1
\end{pmatrix},
\\
\op{\varpi}_{4}
&=
\frac{1}{\sqrt{2}}
\begin{pmatrix}
1
&
- \frac{z}{x} - \frac{1}{2} \frac{\Delta}{x}
\\
- \frac{z}{x} + \frac{1}{2} \frac{\Delta}{x}
&
-1
\end{pmatrix}.
\end{align}
Further, we may normalize the left and right eigenvectors such that
they form a complete and orthonormal basis,
\be
\bbrackket{L_{n}}{R_{m}}=\delta_{nm},
\qquad
\sum_{n} \kket{R_{n}}\bbra{L_{n}}=\op{1},
\ee
and thus we have
\begin{align}
\kket{R_{1}}
&=
\begin{pmatrix}
\frac{1}{\sqrt{2}} \\
-\sqrt{2}\frac{x}{\Delta} \tanh(\frac{\beta\Delta}{2}) \\
0 \\
-\sqrt{2}\frac{z}{\Delta} \tanh(\frac{\beta\Delta}{2})
\end{pmatrix},
\qquad
\kket{R_{2}}
=
\frac{2 z}{\Delta}
\begin{pmatrix}
0 \\
\frac{x}{z} \\
0 \\
1
\end{pmatrix},
\nonumber \\
\kket{R_{3}}
&=
\sqrt{2}\frac{x}{\Delta}
\begin{pmatrix}
0 \\
-\frac{z}{x} \\
-i \frac{1}{2} \frac{\Delta}{x} \\
1
\end{pmatrix},
\qquad
\kket{R_{4}}
=
\sqrt{2}\frac{x}{\Delta}
\begin{pmatrix}
0 \\
-\frac{z}{x} \\
i \frac{1}{2} \frac{\Delta}{x} \\
1
\end{pmatrix}
\nonumber \\
\bbra{L_{1}}
&=
\begin{pmatrix}
\sqrt{2}, &
0, &
0, &
0
\end{pmatrix},
\nonumber \\
\bbra{L_{2}}
&=
\frac{2 z}{\Delta}
\begin{pmatrix}
\frac{1}{2} \Delta \frac{1}{z} \tanh(\frac{\Delta}{2}\beta), &
\frac{x}{z}, &
0, &
1
\end{pmatrix},
\nonumber \\
\bbra{L_{3}}
&=
\sqrt{2}\frac{x}{\Delta}
\begin{pmatrix}
0, &
-\frac{z}{x}, &
i \frac{1}{2} \Delta \frac{1}{x}, &
1
\end{pmatrix},
\nonumber \\
\bbra{L_{4}}
&=
\sqrt{2}\frac{x}{\Delta}
\begin{pmatrix}
0, &
-\frac{z}{x}, &
-i \frac{1}{2} \frac{\Delta}{x}, &
1
\end{pmatrix}.
\end{align}

\section{Lindbladian master equation}
\label{appendix:D}

The equation describing the time evolution of the reduced density
matrix $\op{\rho}$, is a linear and time-local master equation, given
by
\be
\label{eq:lindbladmeq}
\pd_{t} \op{\rho}
=
\widehat{\mathcal{L}}(t) \op{\rho},
\ee
where $\widehat{\mathcal{L}}(t)$ is the Liouvillian written in
Lindblad form (see main text).
In the basis
$\{\op{\Gamma}_{i}\}_{i=1}^{4}
=
\frac{1}{\sqrt{2}}\{\op{1},\op{\sigma}^{x},\op{\sigma}^{y},\op{\sigma}^{z}\}$,
where the vector representation of $\op{\rho}(t)$ reads
$\kket{\rho(t)}=\sum_{i=1}^{4}c_{i}(t)\kket{\Gamma_{i}}$,
Eq.~(\ref{eq:lindbladmeq}) takes the form
\begin{align}
\pd_{t} c_{1}(t)
&=
0
,
\\
\pd_{t} c_{2}(t)
&=
- 4 \pi g^{2} x c_{1}(t)
- 2 \pi g^{2} \coth[\beta b(t)] \frac{x^{2}+b^{2}(t)}{b(t)} c_{2}(t)
\nonumber \\
&\phantom{=}- 2 z(t) c_{3}(t)
- 2 \pi g^{2} \coth[\beta b(t)] \frac{x\,z(t)}{b(t)} c_{4}(t),
\\
\pd_{t} c_{3}(t)
&=
2 z(t) c_{2}(t)
- 2 \pi g^{2} \coth[\beta b(t)] b(t) c_{3}(t)
- 2 x c_{4}(t),
\\
\pd_{t}c_{4}(t)
&=
- 4 \pi g^{2} z(t) c_{1}(t)
- 2 \pi g^{2} \coth[\beta b(t)] \frac{x\,z(t)}{b(t)}
c_{2}(t)
\nonumber \\
&\phantom{=}
+ 2 x\,c_{3}(t)
- 2 \pi g^{2} \coth[\beta b(t)] \frac{b^{2}(t)+z^{2}(t)}{b(t)}
c_{4}(t),
\end{align}
with $b(t)\equiv\sqrt{x^{2}+z^{2}(t)}$.
We numerically solved the above equations to find $\kket{\rho(t)}$.

\end{appendix}



\end{document}